\documentclass[journal,transmag]{IEEEtran}
\usepackage{amsmath,amssymb}
\usepackage{graphicx}
\usepackage{booktabs,multirow}
\usepackage{siunitx}
\usepackage{longtable,array,calc,adjustbox}
\usepackage{upgreek}
\usepackage{enumitem}
\usepackage{comment}
\usepackage{subcaption} 
\usepackage{url}
\usepackage[hidelinks]{hyperref}
\usepackage{cite}

\newcommand{\fitmath}[1]{\begin{adjustbox}{max width=0.94\linewidth}$\displaystyle #1$\end{adjustbox}}

\begin{document}
\title{A Multiobjective Optimization Framework for Sensor Array Design in Magnetocardiography}
\author{
\IEEEauthorblockN{
Kesavaraja C, Sammita Kashyap, Korak Biswas, Pranab Dutta\textsuperscript{*}, and Kushal Patel
}
\IEEEauthorblockA{
GDQLABS Private Limited, Bengaluru, India
}
\thanks{\textsuperscript{*} Corresponding author: Pranab Dutta (e-mail: pranab@gdqlabs.com).}
}
\maketitle
\begin{abstract}
Magnetocardiography (MCG) noninvasively measures cardiac magnetic fields. Conventional cryogenic systems use fixed multichannel arrays, whereas room-temperature sensors enable flexible placement to reduce array size and measurement redundancy. The present study aims at developing a scheme to identify the optimal sensor placements by utilizing three methodological sensor optimization frameworks such as physics driven, experimental data driven and a hybrid combination of both. The physics-driven method optimizes continuous planar coordinates using D-optimality, spatial correlation, and least-squares error. The data-driven method uses measured spatiotemporal fields and pole-geometry metrics, while the hybrid method integrates both approaches. The evaluation is performed using training objective, Pearson correlation, normalized root-mean-square error (NRMSE) and pole separation(mm), orientation(°) and trajectory(mm) errors,  where physics-driven optimization identified 16 sensors with a correlation of 0.9914 and an NRMSE of 0.0139, with pole separation, orientation, and trajectory errors of 3.71 mm, 1.38°, and 3.55 mm, respectively. Data-driven optimization shows diminishing improvements beyond 17 sensors, achieving correlation 0.9840, NRMSE 0.0035, and pole errors of approximately 4.30 mm, 1.70°, and 6.80 mm. The hybrid method achieves correlation 0.9945 and NRMSE 0.0201 with 14 sensors, reducing sensor count by 12.5\% relative to the physics-driven configuration. Its pole errors of 6.50 mm, 2.31°, and 7.50 mm satisfy the study-defined thresholds, balancing array compactness with magnetic-field morphology preservation. These results show that optimized placement helps alleviate the complexity of data redundancy and enhances the performance of MCG as a routine cost-effective clinical measurement.
\end{abstract}
\section{Introduction}\label{introduction}
Magnetocardiography (MCG) records the weak magnetic fields produced by the electrical activity of the heart in a contactless manner~\cite{ref1}. It has shown promising sensitivity for the early detection of cardiac abnormalities, as reported in the literature~\cite{ref2,ref3}. Since MCG signals are inherently weak approximately in the order of 50 pT, their detection requires highly sensitive magnetic sensors such as superconducting quantum interference device sensors, optically pumped magnetometer (OPMs), flux gate sensors, nitrogen - vacancy centers in diamond and magnetoelectric sensors, together with effective suppression of ambient magnetic noise. Among these technologies, OPMs have gained increasing attention for MCG applications because they operate at room temperature, eliminating the need for cryogenic cooling, and allow flexible sensor configurations~\cite{ref4,ref5,ref6}. Moreover, scalar OPMs employed in gradiometric configurations have demonstrated MCG measurements in unshielded environments by suppressing common-mode ambient magnetic interference, highlighting their potential for developing more accessible and portable MCG systems~\cite{ref7,ref8}. 
The simultaneous acquisition of MCG signals at multiple spatial locations of the heart provides the spatial temporal distribution of the magnetic fields at different instants of the cardiac cycle which can be comprised as magnetic field map (MFM) for that specific instant~\cite{ref9,ref10}. However, the number of channels directly increases system cost, data volume, and hardware complexity, while closely spaced channels record strongly correlated, partly redundant information. This redundancy can be eliminated by identifying the optimal sensor positions and then recording the MCG signal at only those chosen positions~\cite{ref11,ref12}. Selecting the optimal sensor count and positions that preserve the essential information of a dense array is therefore a central design problem, particularly for portable unshielded OPM systems that could operate with few channels~\cite{ref7,ref8,ref13}.
The basic problem of optimizing the sensor array for MCG has been explored by many researchers~\cite{ref11,ref12,ref14,ref15,ref16,ref17,ref18,ref19,ref20}. Two families of approaches to this problem have been reported. Physics based methods represent cardiac electrical activity, can be approximated by a finite set of dipole sources and relate them to the sensor signals through the lead field matrix (LFM), computed from a quasi-static forward model. The sensor array is then selected to improve the conditioning of the LFM quantified by condition number (CN) metric compared to other possible sensor array configurations. Lau et al. have used Tabu search, and particle swarm optimization algorithm to identify 20 – 30 vectorial sensors which have the better CN than regular arrays~\cite{ref11} and later designed a 32-sensor flexible vest array covering the front and back of the torso using CN, the Skeel condition number, and the mean singular value as criteria~\cite{ref14}.  Warda et al. compared CN, the singular-value ratio, and the inverse of the squared singular values, and found the singular-value ratio to be the most informative measure of LFM information content~\cite{ref15}. 
Apart from estimating the redundancy of the LFM, the method of solving the inverse problem by finding the fixed predefined dipoles with the estimated source information is also used to choose the optimal sensor array. Van Veen et al. proposed sensitivity to localization error as the cost function of a greedy algorithm, obtaining lower localization root-mean-square error (RMSE) than uniform arrays~\cite{ref16}. Zhang et al. used the uncertainty radius derived from the Cramér–Rao bound and showed that optimized non-uniform arrays of 15 and 25 sensors matched uniform radial-component arrays of 32 and 64 sensors~\cite{ref17}. He et al. combined the Frobenius norm of the LFM with source deviation in a bi-objective cost and found the best results for 48 of 64 candidate channels~\cite{ref18}. Furthermore, Jazbinsek et al. showed that about 20 sites out of 128 sufficed to localize single pre-excitation sites~\cite{ref12}. Although these methods are effective in identifying optimal sensor positions, their performance depends on the assumed forward model, the inverse algorithm, the number, positions, and orientations of the dipole sources, and the quality of the simulated or real MCG signals.
Data-driven approaches identify an optimal reduced sensor configuration by evaluating the extent to which measurements from the selected sensors can reconstruct signals from the non-selected sensors, thereby assessing the information preserved relative to the full reference array. The measured data can be divided at the subject level into training and test sets, with the training set used for sensor selection and the test set reserved for evaluating the performance of chosen sensor array. Previous studies have demonstrated that substantially reduced sensor arrays can reproduce magnetic field maps and selected quantitative features with limited deviations from their full-array references. Using these kind of data driven approaches V. Jazbinsek et al and Kesavaraja et al reported that 20 and 18 channels, selected from 128 and 37 channel recordings, respectively, reproduced the MFM and its quantitative features with minimal deviation from the reference~\cite{ref19,ref20}. 
Existing approaches have the limitation of optimizing complementary and potentially conflicting objectives independently, by considering either source-space metrics, such as localization error and LFM conditioning, or data-driven metrics derived exclusively from experimentally measured signals. An array that is well conditioned for source observability is not necessarily the one that best reproduces measured MFM features, and simulated signals may not capture the full behaviour of real recordings. Moreover, the data driven MFM based metrics such as pole angle, amplitude ratio and trajectory are majorly used for clinical diagnostics such as coronary artery disease, hypertrophy and pulmonary hypertension rather than estimating the cardiac source of these anomalies by modelling~\cite{ref2,ref3,ref9}. Although physics-based sensor selection and data-driven field-map reconstruction have both been investigated, their integration within a single continuous-coordinate optimization framework using the present combination of source-information and MFM-reconstruction criteria remains underexplored.
This paper addresses this gap by determining the sensor count N and sensor positions \((u,v)\) that retain the useful information of a 32-channel reference array through a hybrid physics-driven and data-driven optimization. In the physics-driven part, MCG signals are simulated from ventricular action potential propagation in a realistic torso model with piecewise-homogeneous conductivities for each organs. The resulting LFM and MFMs at selected instants of the cardiac cycle are used to evaluate the cardiac source information captured by different sensor configurations. In the data-driven part, MFMs reconstructed from the reduced optimized array are compared with those of the experimentally measured 32-channel MCG data using pole orientation, pole-distance, correlation, and associated metrics. By integrating these complementary approaches into a unified cost function, the proposed framework seeks to optimize sensor configurations that preserve the simulated cardiac source information while maintaining the spatial characteristics of experimentally measured MCG signals. The main contributions of the present paper are as follows:
\begin{enumerate}[label=(\roman*)]
\item {A combined cost function which integrates LFM-based cardiac source information with the reconstruction accuracy of experimentally measured MFMs.}
\item{A systematic evaluation of the dependence of reconstruction accuracy and spatial coverage on sensor count across physics-driven, data-driven, and hybrid optimization formulations.}
\end{enumerate}
The remainder of this paper is organized as follows. Section ~\ref{Materials-and-methods} presents the cardiac source model, forward field computation and LFM extraction, followed by the data driven reconstruction and the combined objective formulation. Section~\ref{Results} reports the results for each method and their comparison. Section ~\ref{Discussion}  presents the reason and significance of the obtained results including count behaviour, reconstruction accuracy and spatial coverage, for each method separately and compare the complementary strengths of the frameworks in the combined approach and the limitations of the present work. Section ~\ref{Conclusion} concludes with the main findings, and directions for further validation.
\section{Materials and methods}\label{Materials-and-methods}
\subsection{Physics driven optimization}\label{physics-driven-optimization}
In the physics framework, cardiac ventricular activity and its resulting magnetic fields are simulated using cardiac source dipoles with three Cartesian dipole components for the design of optimized sensor array. The electrical activity of ventricular myocardium is modeled using finite element method (FEM) within a realistic human anatomical model. The resulting cardiac magnetic fields are then calculated at the sensor positions using the Biot–Savart law through the LFM~\cite{ref21}.
\begin{figure*}[!t]
    \centering
    \begin{subfigure}[t]{0.31\textwidth}
        \centering
        \includegraphics[width=\linewidth]{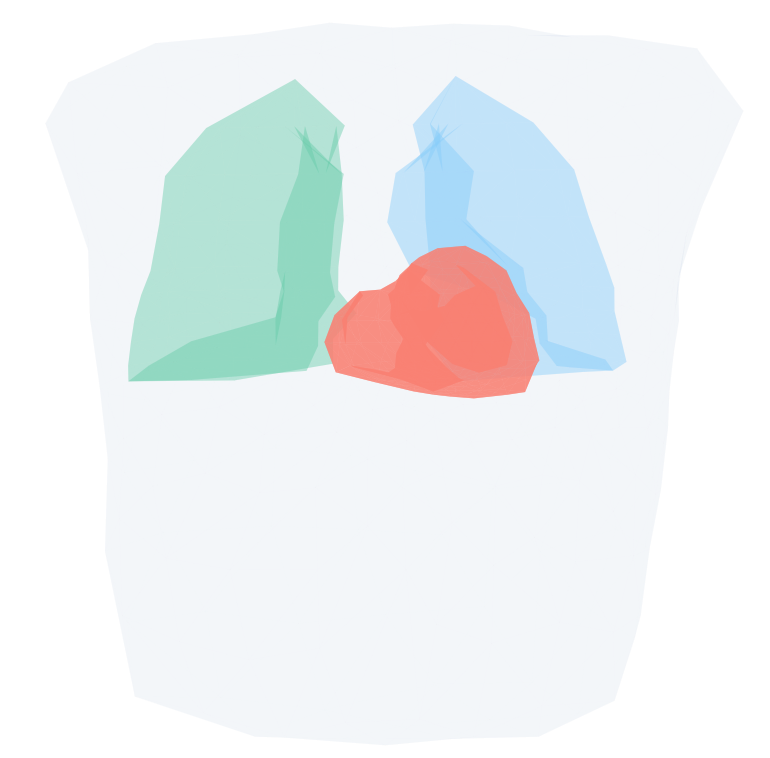}
        \caption{}
        \label{fig:source_model_torso_heart}
    \end{subfigure}\hfill%
    \begin{subfigure}[t]{0.31\textwidth}
        \centering
        \includegraphics[width=\linewidth]{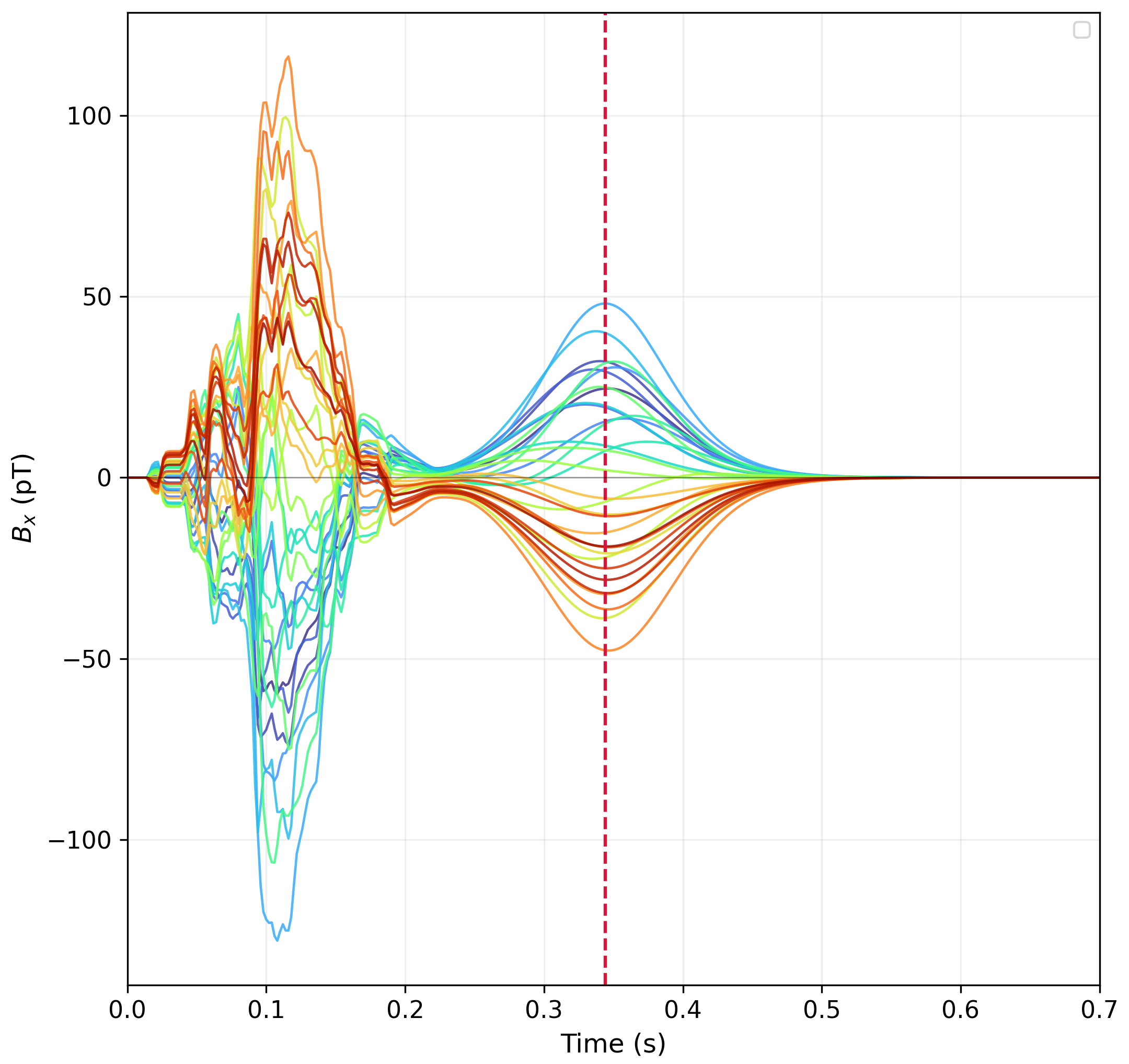}
        \caption{}
        \label{fig:source_model_magnetic_signal_T_wave}
    \end{subfigure}\hfill%
    \begin{subfigure}[t]{0.38\textwidth}
        \centering
        \includegraphics[width=\linewidth]{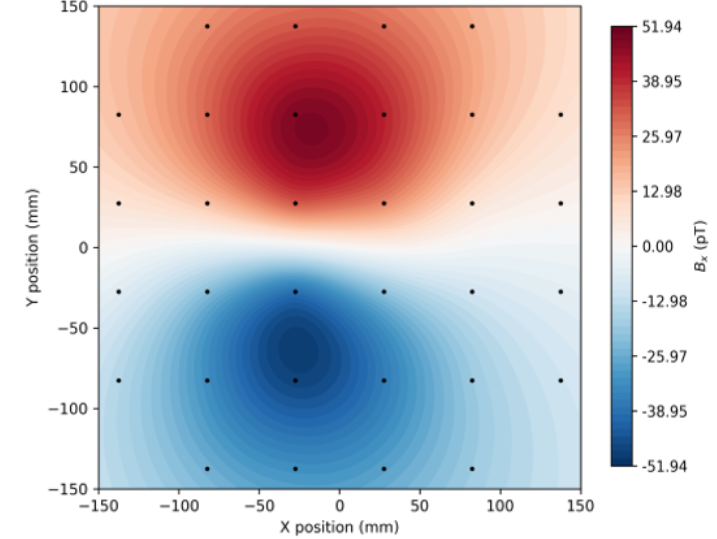}
        \caption{}
        \label{fig:source_model_32_ref_mfm}
    \end{subfigure}
    \caption{Generating magnetic signals using simulated source model of the heart:
    (a) Cardiac source model,
    (b) MCG signals generated at 32 sensor positions, and
    (c) MFM obtained at the T-wave peak instant.}
    \label{fig:MCG-signal-simulation}
\end{figure*}
\subsubsection{MCG signal simulation}\label{MCG-signal-simulation}
The cardiac electrical activity is represented by time-varying equivalent current dipoles at 257 fixed locations on the heart surface geometry and the  activation propagation defined by action potential is taken from ECGSIM covering the ventricular depolarization, repolarization times~\cite{ref22,ref23,ref24}. Each dipole can have three free orientations, so the complete state contains 771 coefficients. A current dipole is a compact representation of a localized current distribution; its direction specifies current orientation and its magnitude gives the dipole moment in \(\mathbf{A\,m}\). The source positions define where activity can occur, while the coefficient trajectories define how its strength and orientation evolve. With source location \(\mathbf r_j\) and moment \(\mathbf p_j(t)\), the impressed-current density representation is
\begin{equation}
\label{eq:2}
\fitmath{\mathbf J_p(\mathbf r,t)=\sum_{j=1}^{N_s}\mathbf p_j(t)\,\delta(\mathbf r-\mathbf r_j),\qquad \mathbf p(t)=[p_{1x}(t),p_{1y}(t),p_{1z}(t),\ldots,p_{257z}(t)]^T}
\end{equation}
Here, \(N_s=257\) is the number of source nodes, \(\mathbf J_p(\mathbf r,t)\) is the impressed-current density at position \(\mathbf r\) and time \(t\), and \(\delta(\cdot)\) is the three-dimensional Dirac delta distribution, which localizes each dipole at \(\mathbf r_j\). The superscript \(T\) denotes transpose. The source matrix \(\mathbf P\in\mathbb R^{T\times771}\) contains \(T=351\) sampled states spanning QRS and T waves, with each row equal to the transpose of the corresponding source vector \(\mathbf p(t)\). Columns are ordered as the x, y and z components for each source node. Hereafter, \(N\) denotes the number of candidate sensors, whereas \(N_s\) denotes the fixed number of source nodes. The same state drives the reference and candidate arrays, so differences between their predicted fields arise from sensor geometry and reconstruction rather than a change of cardiac excitation. Linear superposition combines all nodal contributions, including reinforcement and cancellation between differently oriented sources.
The heart sources are embedded in a tetrahedral volume conductor whose material tags assign a scalar, piecewise-constant conductivity. The configured conductivities are 0.20 S/m for torso and myocardium, 0.05 S/m for each lung, and 0.60 S/m for the cardiac blood cavities. The mesh contains 93,043 nodes and 543,350 elements for all the organs. The supplied labels include torso, right lungs, left lungs, heart ventricles, the right-cavity and left-cavity-tagged elements are present. The entire ventricular activity regime is used in the magnetic field reconstruction objectives~\cite{ref21}.
The forward model maps a prescribed cardiac current distribution to the magnetic field sampled by an array. At each time point, the response is the sum of the fields produced by all source components. The conductivity geometry and sensor locations determine the transfer relationship, while the source moments determine its time-varying excitation~\cite{ref1}. The measured component is the projection of the total field onto the sensor sensing axis, normal to the sensor plane.
For a dipole at \(\mathbf r_j\), the direct field at sensor position \(\mathbf r_i\) is computed from the current-dipole Biot--Savart expression,
\begin{equation}
\label{eq:4}
\fitmath{\mathbf B_0(\mathbf r_i;\mathbf r_j,\mathbf p_j)=\frac{\mu_0}{4\pi}\frac{\mathbf p_j\times(\mathbf r_i-\mathbf r_j)}{\|\mathbf r_i-\mathbf r_j\|^3}}
\end{equation}
Here, \(\mu_0\) is the permeability of free space, \(\times\) denotes the vector cross product, and \(\|\cdot\|\) denotes the Euclidean norm. The vector \(\mathbf r_i-\mathbf r_j\) points from source \(j\) to sensor \(i\). Positions are expressed in metres for this calculation, giving the primary magnetic field \(\mathbf B_0\) in tesla.

Conductivity boundaries redistribute return currents, so the total field also includes a potential-dependent interface contribution. The implementation extracts triangular faces shared by two tetrahedra with different conductivities, orients their normals consistently with the conductivity jump, and interpolates the nodal potentials at quadrature points. Its evaluation uses a centroid rule on each interface triangle:
\begin{equation}
\label{eq:5}
\fitmath{\mathbf B(\mathbf r_i)=\mathbf B_0(\mathbf r_i)-\frac{\mu_0}{4\pi}\sum_q\Delta\sigma_q\,\phi_q\,\frac{(\mathbf r_i-\mathbf r_q)\times\mathbf n_q\,w_q}{\|\mathbf r_i-\mathbf r_q\|^3}}
\end{equation}
The index \(q\) runs over the interface quadrature points, and \(\phi_q\) is the electric potential, in volts, evaluated at \(\mathbf r_q\). With one centroid quadrature point per triangle, \(w_q\) equals the triangle area. For the complete source distribution, \(\mathbf B_0(\mathbf r_i)\) denotes the sum of the primary fields from all source dipoles, whereas \(\mathbf B(\mathbf r_i)\) includes both primary and conductivity-interface contributions.

In this expression \(\mathbf r_q\) is an interface quadrature point, \(\mathbf n_q\) is its oriented unit normal, \(w_q\) is the area weight and \(\Delta\sigma_q\) is the conductivity difference across that face. Geselowitz's formulation provides the physical background for this potential-based treatment of conductivity interfaces~\cite{ref21}. The heart - torso model, the resulting simulated MCG signals and the MFM obtained at T peak instant are displayed in Figure~\ref{fig:MCG-signal-simulation}.
\subsubsection{Sensor array optimization}\label{sensor-array-optimization}
 The decision vector contains two positional coordinates per sensor:
\begin{equation}
\label{eq:8}
\fitmath{\mathbf x=[u_1,v_1,u_2,v_2,\ldots,u_N,v_N]^T}
\end{equation}
The decision vector \(\mathbf x\in\mathbb R^{2N}\) contains the two in-plane coordinates of each of the \(N\) candidate sensors, expressed in millimetres.

Here, \(u_i\) and \(v_i\) are positions along the original anatomical
\(y\) and \(z\) directions relative to the whole-heart centroid.
All sensors lie on the frontal plane
\(x=x_{\mathrm{centroid}}+100\,\mathrm{mm}\), with lateral limits
of \(\pm135\,\mathrm{mm}\). The geometry constrains such as all sensor pairs must remain at least \(54\,\mathrm{mm}\)
apart, and each sensor must have a neighbor within
\(65\,\mathrm{mm}\) are kept to match with the practical OPM sensor systems. 
The same geometry and plane-normal component are used for
D-optimality, Pearson correlation, and least-squares reconstruction.
A \(5\,\mathrm{mm}\) grid covering the
\(270\times270\,\mathrm{mm}\) sensor plane contains
\(55\times55\) positions, each with 771 lead-field coefficients. Multiplication by the source matrix with LFM
provides magnetic field values for all time instants. 
\subsubsection{Cost function}\label{Cost-function}
The objective function for physics-driven sensor-array optimization comprises three distinct terms and is defined as follows:
\begin{equation}
\label{eq:16}
\fitmath{
J_{\mathrm{physics}}
=
w_{D}\frac{J_D}{s_D}
+
w_{P}\,J_P
+
w_{LS}\,\frac{J_{LS}}{s_{LS}}
}
\end{equation}
The fixed positive normalization scales are
\[
s_D=\max\!\left(\left|J_D^{(16)}\right|,10^{-12}\right),
\qquad
s_{LS}=\max\!\left(J_{LS}^{(16)},10^{-12}\right).
\]
The superscript \((16)\) denotes evaluation at the centered 16-sensor anchor layout using the same lookup model and complete source sequence. The lower bound prevents division by zero.

where \(J_{D}\), \(J_{P}\) and \(J_{LS}\) denote D-optimality, Pearson correlation and least square(LS) error, respectively, with fixed weights for each 0.15, 0.70 and 0.15. The D-optimality and LS terms are normalized using fixed reference values obtained from a centered 16-sensor baseline layout. These normalization constants are retained for all sensor counts, ensuring a consistent objective scale throughout the sweep.
Consequently, the prior, kernel parameters, and objective scales
are common to the array comparison. 
\begin{enumerate}[label=(\roman*), leftmargin=1.5em,labelsep=0.4em]
\item {D-optimality:}
It is an information-based design criterion used to evaluate sensor configurations according to their ability to capture information about the underlying sources. In the present study, it is formulated using the FEM - LFM and the assumed source, sensor-noise covariance models. The resulting regularized sensor-space matrix represents the information captured by the sensor configuration, while its log-determinant provides a scalar measure of the overall information content. Accordingly, minimizing the negative log-determinant favors sensor configurations with greater information content under the assumed statistical model~\cite{ref25}.
Let \(\mathbf L_N\in\mathbb R^{N\times771}\) denote the candidate
FEM lead-field matrix. Assuming an isotropic source covariance
\(\sigma_q^2\mathbf I_{771}\) and independent sensor-noise covariance
\(\sigma_n^2\mathbf I_N\), the regularized sensor-space matrix and
minimized design term are
\begin{equation}
\label{eq:9}
\fitmath{
g=(\sigma_q/\sigma_n)^2,\qquad
\mathbf F=\mathbf I_N+g\mathbf L_N\mathbf L_N^T,\qquad
J_D=-\log\det\mathbf F
}
\end{equation}
Here, \(\mathbf I_m\) denotes the \(m\times m\) identity matrix, and \(g\) is the ratio of source-prior variance to sensor-noise variance. The matrix \(\mathbf F\in\mathbb R^{N\times N}\) is the sensor-space covariance normalized by the noise variance. The operators \(\det\) and \(\log\) denote the determinant and natural logarithm, respectively.

The fixed parameters are the prior standard deviation of the dipole moment amplitude,
\(\sigma_q=10^{-7}\,\mathrm{A\,m}\), and the sensor noise standard deviation,
\(\sigma_n=1.5\times10^{-14}\,\mathrm T\) (15 fT), assumed for shielded conditions.
Lower \(J_D\) rewards larger
regularized determinant and complementary source sensitivities~\cite{ref25}.
\item {Pearson correlation:}
Pearson correlation quantifies the similarity between the spatial
distributions of the reconstructed candidate MFMs and the corresponding 32-channel reference MFMs~\cite{ref20}. The source matrix \(\mathbf P\in\mathbb R^{T\times771}\) contains successive dipole states, while \(\mathbf L_{32}\) and \(\mathbf L_N\) represent the lead-field matrices for the reference and candidate geometries. The simulated signals are arranged as
\(\mathbf B_{32}\in\mathbb R^{32\times T}\) and
\(\mathbf B_N\in\mathbb R^{N\times T}\):
\begin{equation}
\label{eq:10}
\fitmath{
\mathbf B_{32}=\mathbf L_{32}\mathbf P^T,\qquad
\mathbf B_N=\mathbf L_N\mathbf P^T.
}
\end{equation}
To compare the spatial field distributions at identical locations,
the candidate measurements are reconstructed at the fixed reference
sensor positions using Gaussian kernel ridge regression
\cite{ref26}. The spatial kernel and regularized regression
system are defined in \eqref{eq:12}, with kernel length scale
\(\ell=60\,\mathrm{mm}\) and regularization parameter \(\lambda=0.03\):
\begin{equation}
\label{eq:12}
\fitmath{
\begin{gathered}
K(\mathbf{s},\mathbf{s}')
=\exp\!\left[-\frac{\|\mathbf{s}-\mathbf{s}'\|^2}{2\ell^2}\right],
\\[6pt]
(\mathbf{K}_{ss}+\lambda\mathbf{I}_N)\boldsymbol{\alpha}
=\mathbf{B}_N,\qquad
\hat{\mathbf{Y}}_K=\mathbf{K}_{rs}\boldsymbol{\alpha}.
\end{gathered}
}
\end{equation}
For candidate positions \(\mathbf s_i\) and reference positions \(\mathbf s_a^{\mathrm{ref}}\),
\[
[\mathbf K_{ss}]_{ij}=K(\mathbf s_i,\mathbf s_j),\qquad
[\mathbf K_{rs}]_{ai}=K(\mathbf s_a^{\mathrm{ref}},\mathbf s_i).
\]
Thus, \(\mathbf K_{ss}\in\mathbb R^{N\times N}\) and \(\mathbf K_{rs}\in\mathbb R^{32\times N}\). Both \(\mathbf s\) and \(\mathbf s'\) are in-plane position vectors. The matrix \(\hat{\mathbf Y}_K\in\mathbb R^{32\times T}\) contains the kernel-reconstructed signals at the reference sensor positions.

Here, \(\mathbf s=(u,v)^T\),
\(\mathbf K_{ss}\) relates candidate sensor positions,
\(\mathbf K_{rs}\) relates reference to candidate positions, and
\(\boldsymbol{\alpha}\in\mathbb R^{N\times T}\) contains the
regression coefficients. The reference field matrix is
\(\mathbf Y=\mathbf B_{32}\).
For two nonconstant vectors, \(\rho(\mathbf a,\mathbf b)\)
denotes their Pearson correlation after subtracting their respective
means. At each sample, spatial correlation compares the reconstructed
and reference magnetic-field values across the 32 reference positions.
The Pearson objective is the equally weighted mean correlation penalty
across valid reference samples:
\begin{equation}
\label{eq:13}
\fitmath{
\begin{gathered}
r_s(t)=
\rho\!\left(\hat{\mathbf Y}_{K,:,t},\mathbf Y_{:,t}\right),
\\
J_P=
1-\frac{1}{|\mathcal T_s|}
\sum_{t\in\mathcal T_s}\tilde r_s(t).
\end{gathered}
}
\end{equation}
The notation \(\mathbf Y_{:,t}\) selects all reference positions at sample \(t\), and \(|\mathcal T_s|\) is the number of samples in the valid-reference set.

The set \(\mathcal T_s\) contains samples with non constant reference
field maps, and \(\tilde r_s(t)\) denotes the spatial correlation
after applying the implementation's handling of degenerate
reconstructions. Each valid reference sample receives equal weight.
Higher spatial correlation indicates closer agreement in the
signed field pattern, while minimizing \(J_P\) promotes preservation
of this pattern across the evaluated samples.
\item {Least square error:}
The least-square (LS) criterion quantifies the amplitude-sensitive reconstruction error between the reference magnetic field signals and its reconstructed counterpart. For each source sample, the regularized source estimate and the corresponding reconstructed reference response are computed as follows:
\begin{equation}
\label{eq:14}
\fitmath{
\hat{\mathbf Q}
=g\mathbf L_N^T\mathbf F^{-1}\mathbf B_N,
\qquad
\hat{\mathbf Y}_{LS}
=\mathbf L_{32}\hat{\mathbf Q}
}
\end{equation}
The matrix \(\hat{\mathbf Q}\in\mathbb R^{771\times T}\) contains the estimated Cartesian dipole moments for all source nodes and samples. The matrix \(\hat{\mathbf Y}_{LS}\in\mathbb R^{32\times T}\) contains the corresponding reconstructed magnetic signals at the fixed reference positions. The subscript \(LS\) distinguishes this source-based reconstruction from the kernel reconstruction \(\hat{\mathbf Y}_K\).

This estimate corresponds to the posterior mean under the assumed isotropic source prior and sensor-noise model~\cite{ref26}. The associated linear systems are solved using Cholesky factorization rather than explicit matrix inversion~\cite{ref27}.
A single reference root mean square (RMS) over all channels and samples defines
the normalization:
\begin{equation}
\label{eq:15}
\fitmath{
\begin{gathered}
h=
\sqrt{\frac{1}{32T}
\sum_{t=0}^{T-1}\sum_{i=1}^{32}Y_{it}^{\,2}},
\\
J_{LS}=
\frac{\|\hat{\mathbf Y}_{LS}-\mathbf Y\|_F^2}
     {32T h^2},
\qquad
\mathrm{NRMSE}_{LS}
=\sqrt{J_{LS}}
=\frac{\|\hat{\mathbf Y}_{LS}-\mathbf Y\|_F}
       {\|\mathbf Y\|_F}
\end{gathered}
}
\end{equation}
Here, \(Y_{it}\) is the reference magnetic-field value at sensor \(i\) and sample \(t\). The Frobenius norm is defined by~\cite{ref27} 
\[
\|\mathbf A\|_F=\sqrt{\sum_i\sum_t A_{it}^{\,2}}.
\]
The quantity \(h\) has the same units as the field, whereas \(J_{LS}\) and \(\mathrm{NRMSE}_{LS}\) are dimensionless. The normalization requires a nonzero reference signal.

The resulting LS objective and normalized root-mean-square error (NRMSE) quantify the discrepancy between the reconstructed and reference field signals. Normalization by the global reference root mean square (RMS) avoids division by individual near-zero samples and expresses the total reconstruction error relative to the overall reference signal strength. It preserves the relative amplitude contributions of different channels rather than normalizing each channel independently. The same whole-sequence NRMSE formulation is applied to the kernel ridge regression reconstruction. At each source sample, the absolute reconstruction error across the 32 reference positions is also reported in pT. Division by the fixed normalization factor \(h\) provides the time-resolved normalized error. These error measures remain available even when the corresponding Pearson correlation is undefined. Thus, the LS criterion complements the correlation-based metric by penalizing amplitude discrepancies in the reconstructed reference signals.
\end{enumerate}
\subsection{Data driven sensor array optimization}\label{data-driven-sensor-array-optimization}
\subsubsection{MCG data}\label{MCG-data}
\begin{figure*}[!t]
    \centering
    \begin{subfigure}[t]{0.38\textwidth}
        \centering
        \includegraphics[width=\linewidth]
        {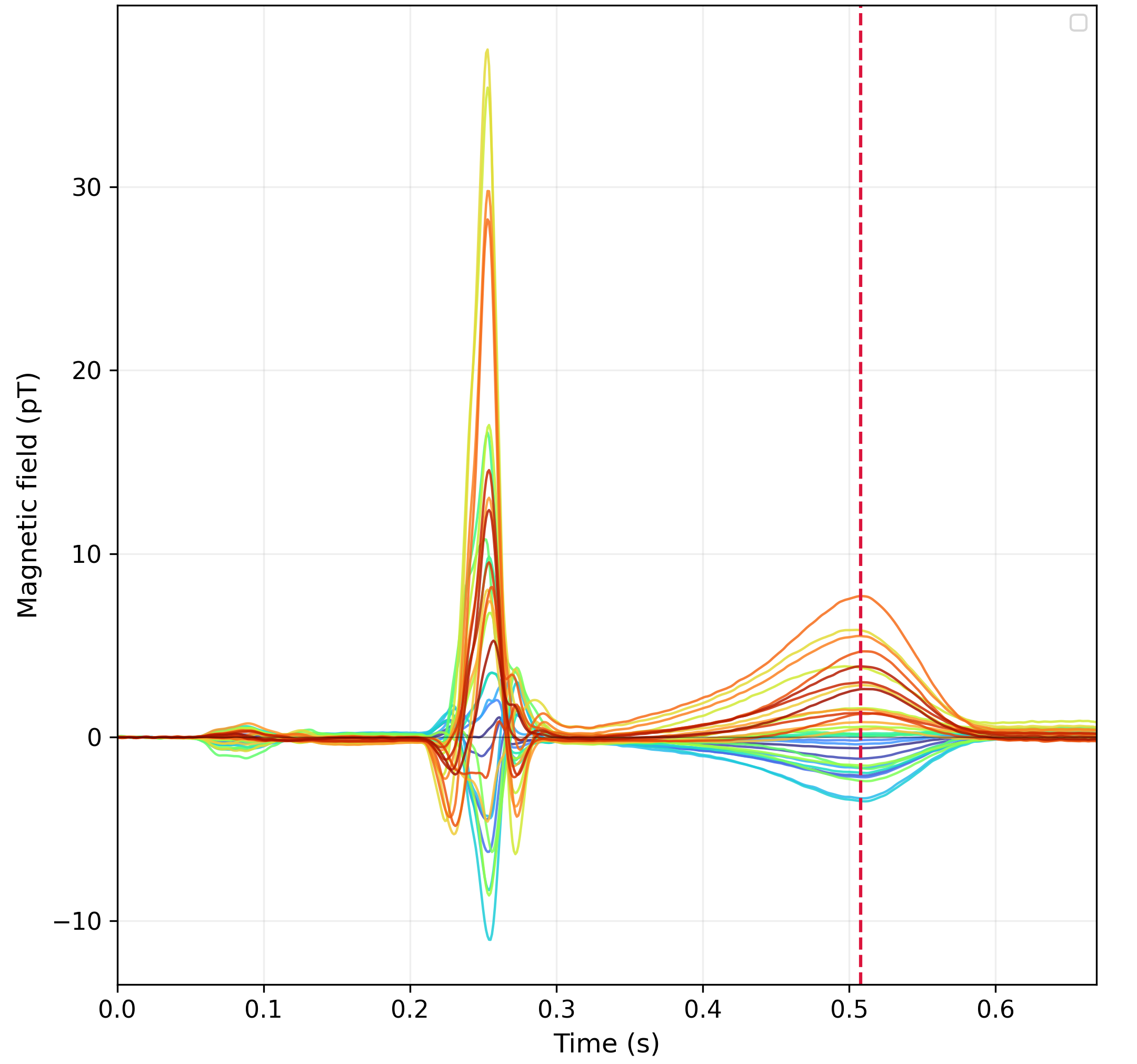}
        \caption{}
        \label{fig:data-t-wave-signal}
    \end{subfigure}
    \begin{subfigure}[t]{0.44\textwidth}
        \centering
        \includegraphics[width=\linewidth]
        {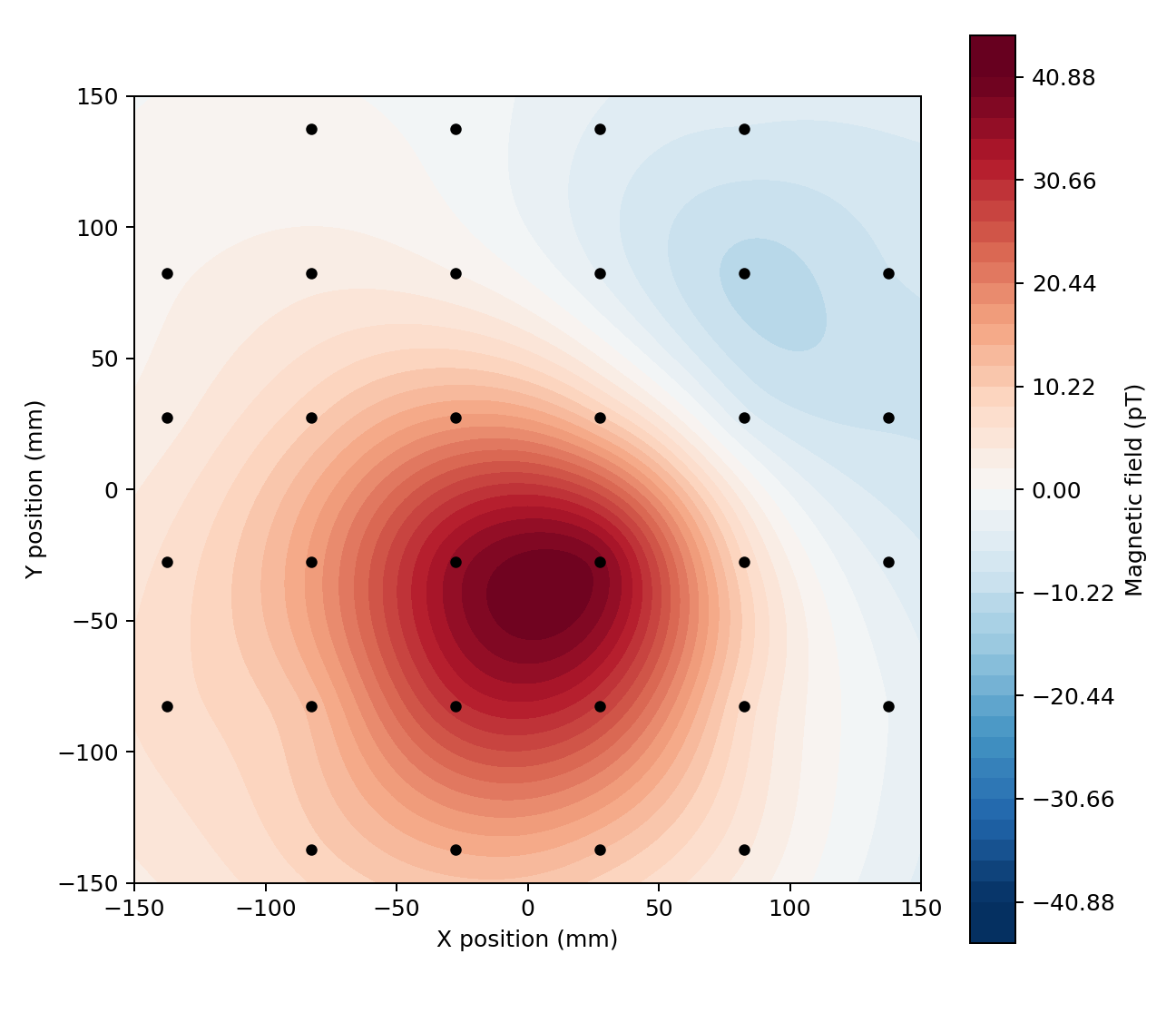}
        \caption{}
        \label{fig:data-driven-reference-mfm}
    \end{subfigure}
    \caption{Real MCG signals of a subject and the corresponding MFM obtained at the T-wave peak instant: (a) signal averaged MCG signals of a single cardiac cycle for 32 channel sensor array; (b) MFM at T peak.}
    \label{fig:data-driven-reference}
\end{figure*}

In the data driven approach, the publicly available MCG dataset hosted on IEEE DataPort~\cite{ref28}  is used for sensor array optimization. The normal component of MCG data is recorded using 32 OPM sensors kept in the frontal plane to the chest in a shielded environment. The MCG signals are preprocessed to remove the 50 Hz, its harmonic noise and low frequency drift. The signals are sampled at 1000 Hz and recorded for a period of 60 seconds with intersensor spacing of 55 mm. The preprocessed signals are temporally averaged using R peak trigger to get the single cardiac cycle for all channels. This averaged signals are interpolated using thin-plate-spline (TPS) interpolant  to get smoothed magnetic field map similar to the case of simulated MCG signals at any desired time instant is shown in Figure~\ref{fig:data-driven-reference}~\cite{ref29}. The present work considers the MCG data of 6 subjects which have clear MFM dipolar pattern. Out of which 3 subject data used for training the optimization algorithm and 3 are used for evaluation of the chosen optimized sensor array. 
For each candidate geometry, the TPS interpolated MFM is evaluated at the candidate sensor locations. The resulting candidate sensor signals are subsequently used to reconstruct the field on a common evaluation grid~\cite{ref29}. This enabled the reconstructed field obtained with \(N\) sensors, denoted by \(M_N(q,t)\), to be compared with the reference field \(M_{32}(q,t)\) at identical spatial and temporal coordinates. The interpolated signals at candidate locations represent estimates derived from the reference field rather than additional physical measurements. Consequently, the optimization identifies candidate sensor configurations based on their ability to preserve the reference field characteristics.
\subsubsection{Cost function}\label{data-driven-cost-function}
The data-driven objective function was formulated to preserve the spatial and temporal characteristics of the experimentally measured MFMs while reducing the number of sensors. The objective evaluates reconstruction fidelity in terms of field amplitude, spatial morphology, positive and negative field-lobe geometry, and temporal consistency. In addition, penalties associated with sensor distribution, measurement redundancy, spatial coverage, and numerical stability are incorporated to discourage configurations that provide good reconstruction accuracy but are poorly distributed or computationally unstable. The evaluation of reduced sensor configurations through the preservation of reference MFMs follows the approach adopted in earlier MCG channel-selection studies~\cite{ref20}. The overall data-driven objective is expressed as
\begin{equation}
\label{eq:22}
\fitmath{\begin{aligned}J_{data}&=w_PE_P+w_AE_A+w_{pole}E_{pole}+w_{\nabla}E_{\nabla}+w_{div}E_{div}+w_{wcov}E_{wcov}\\
 &{}+w_{drop}E_{drop}+w_{sep}E_{sep}+w_{\theta}E_{\theta}+w_{traj}E_{traj}+w_{cov}E_{cov}+w_{cond}E_{cond}\end{aligned}}
\end{equation}
The numerical weights are reported in Section~\ref{Combined-physics-informed-and-data-driven-objective}. These coefficients control the relative contributions of the terms and are not required to sum to unity.

where each (\(E\)) represents an individual error or penalty metric and the corresponding (\(w\)) denotes its weighting coefficient.
\begin{enumerate}[label=(\roman*), nosep, leftmargin=1.5em,labelsep=0.4em]
\item {Correlation error:}
The spatial correlation term, \(E_P\), quantifies the loss of spatial similarity between the reference and reconstructed MFMs using the Pearson correlation coefficient. To account for both the overall reconstruction performance and the preservation of individual high-activity frames, the metric incorporates the mean correlation across all evaluated frames and the minimum correlation among active frames:
\begin{equation}
\label{eq:23}
\fitmath{E_P=(1-\bar r)+\lambda_{worst}(1-r_{min,active})}
\end{equation}
Here, \(\bar r\) denotes the mean Pearson correlation across the evaluated frames, and \(r_{\min,\mathrm{active}}\) represents the minimum correlation among the active frames. The parameter \(\lambda_{\mathrm{worst}}=1.39\) controls the contribution of the worst-frame correlation to the overall error. A frame is classified as active when its peak absolute reference-field amplitude reaches at least \(5\%\) of the maximum reference-field amplitude across the evaluated frames. Frames below this threshold are excluded from the worst-frame correlation calculation but are retained in the mean correlation calculation. This formulation prevents poor reconstruction of individual high-activity frames from being masked by good average performance.
\item {Amplitude error:}
The amplitude reconstruction term, \(E_A\), measures the difference in field magnitude between the reference and reconstructed MFMs. It is calculated as the mean frame-wise root-mean-square error (RMSE), normalized by the maximum absolute reference-field amplitude across all evaluated frames and spatial grid points:
\begin{equation}
\label{eq:23a}
\fitmath{\quad E_A=\frac{1}{T A_{max}}\sum_t\sqrt{\frac{1}{Q}\sum_q[M_N(q,t)-M_{32}(q,t)]^2}}
\end{equation}
The summations cover \(t=0,\ldots,T-1\) and \(q=1,\ldots,Q\), with
\[
A_{\max}=\max_{q,t}|M_{32}(q,t)|.
\]
The amplitude objective \(E_A\) is dimensionless and requires \(A_{\max}>0\). It averages frame-wise RMSE normalized by peak reference amplitude and is therefore distinct from the global energy-normalized NRMSE used for evaluation.

Here, \(M_{N}(q,t)\) and \(M_{32}(q,t)\) denote the reconstructed and reference magnetic field values, respectively, at spatial grid point (q) and time instant (t). The variables (T) and (Q) represent the number of evaluated time frames and spatial grid points, respectively, while \(A_{\max}\) is the maximum absolute reference-field value over the complete evaluation domain. A single global amplitude normalization is used to ensure consistent scaling across all evaluated frames. Consequently, this metric penalizes deviations in field magnitude while maintaining comparability between different sensor configurations.
\item {Pole position error:}
The pole-position term, \(E_{\mathrm{pole}}\), evaluates the spatial displacement of the positive and negative magnetic field-lobe centroids between the reference and reconstructed MFMs~\cite{ref30}. Rather than identifying the pole positions solely from the locations of the maximum and minimum field values, the centroids are calculated using the spatially weighted positive and negative field distributions over the complete evaluation grid:
\begin{equation}
\label{eq:24}
\fitmath{\mathbf c_+(t)=\frac{\sum_q\max(M_q(t),0)\mathbf r_q}{\sum_q\max(M_q(t),0)},\quad \mathbf c_-(t)=\frac{\sum_q\max(-M_q(t),0)\mathbf r_q}{\sum_q\max(-M_q(t),0)}}
\end{equation}
Equation~\eqref{eq:24} is applied separately to \(M_{32}(q,t)\) and \(M_N(q,t)\) to obtain the reference and reconstructed centroids, respectively. The operator \(\max(a,0)\) retains only the positive part of \(a\). A centroid is undefined when its denominator is zero, indicating that the corresponding polarity is absent. Centroid coordinates are expressed in millimetres.

where \(M_q(t)\) denotes the magnetic field value at grid point (q) and time instant (t), and \(\mathbf r_q\) represents its corresponding spatial coordinate. The centroids are calculated independently for the positive and negative field distributions, thereby characterizing the spatial locations of both field lobes. The reference and reconstructed centroids are denoted by \(\mathbf c_{\pm}(t)\) and \(\hat{\mathbf c}_{\pm}(t)\), respectively. Frames in which either field lobe is undefined are handled using the corresponding validity masks implemented in the objective function. This metric encourages the optimized sensor configuration to preserve the spatial locations of the characteristic positive and negative magnetic field regions.
\item {Spatial field-gradient preservation error:}
The field-gradient term, \(E_{\nabla}\), evaluates the difference between the spatial gradients of the reference and reconstructed MFMs~\cite{ref31}. While the amplitude and correlation metrics primarily assess field magnitude and overall spatial similarity, the gradient metric provides an additional measure of local spatial variation. Preserving the field gradients is important for retaining the spatial transitions and local morphological characteristics of the magnetic field distribution~\cite{ref30}. Together with the pole-position metric, it helps ensure that the reconstructed field maps retain both the overall dipolar structure and its local spatial variations.
\item {Pole-pair geometry and trajectory preservation:}
In addition to preserving the individual pole positions, the relative geometry and temporal evolution of the positive and negative field lobes are evaluated using pole-separation, orientation, and trajectory metrics. The reference pole separation, \(d(t)\), and orientation, \(\theta(t)\), are defined as
\begin{equation}
\label{eq:25}
\fitmath{
\begin{aligned}
d(t) &= \lVert \mathbf{c}_{-}(t)-\mathbf{c}_{+}(t)\rVert,\\
\theta(t) &= \operatorname{atan2}
\left(c_{-,y}(t)-c_{+,y}(t),
c_{-,x}(t)-c_{+,x}(t)\right)
\end{aligned}
}
\end{equation}
Here, \(c_{\pm,x}\) and \(c_{\pm,y}\) are the horizontal and vertical coordinates of the pole centroids in the two-dimensional MFM display plane. For the physics geometry, these coordinates correspond to the anatomical lateral coordinates \((u,v)\), rather than the anatomical plane-normal \(x\) coordinate. The function \(\operatorname{atan2}(\Delta y,\Delta x)\) returns the quadrant-aware angle, in radians, measured from the positive horizontal axis toward the positive-to-negative pole direction. The separation \(d(t)\) is expressed in millimetres. The reconstructed quantities \(\hat d(t)\) and \(\hat\theta(t)\) are obtained by replacing the reference centroids with reconstructed centroids.

The separation error, \(E_{\mathrm{sep}}\), quantifies the discrepancy between the reference and reconstructed distances separating the two field-lobe centroids. The orientation error, \(E_{\theta}\), evaluates the difference in the angular orientation of the pole pair. Wrapped angular differences within the interval \([-\pi,\pi]\) are used to account for angular periodicity and avoid discontinuities at the angular boundaries.
The trajectory term, \(E_{\mathrm{traj}}\), evaluates the temporal consistency of the reconstructed pole positions by considering the displacement of both centroids across the evaluated frames. The displacement of each pole centroid at sample \(t\) is expressed as
\begin{equation}
\label{eq:26}
\fitmath{
e_{\pm}(t)=
\left\lVert
\hat{\mathbf c}_{\pm}(t)-\mathbf c_{\pm}(t)
\right\rVert
}
\end{equation}
The symbols \(e_+(t)\) and \(e_-(t)\) denote separate Euclidean position errors for the positive and negative poles, respectively, in millimetres. The hat denotes a quantity obtained from the reconstructed map. In contrast to the worst-frame training penalty, the reported trajectory evaluation error averages position errors over both polarities and all valid timestamps.

The trajectory metric considers the worst-frame displacement to penalize configurations that exhibit substantial pole-position deviations at individual time instants. These complementary metrics assess whether the reduced sensor configuration preserves not only the individual pole locations but also their relative spatial arrangement and temporal evolution throughout the cardiac cycle.
\end{enumerate}
The additional terms \(E_{\mathrm{div}}\), \(E_{\mathrm{cov}}\),
\(E_{\mathrm{wcov}}\), \(E_{\mathrm{drop}}\), and
\(E_{\mathrm{cond}}\) penalize measurement redundancy,
insufficient spatial coverage, insufficient coverage of regions
with greater field variability, sensitivity to individual sensor
removal, and poor conditioning of the TPS reconstruction system,
respectively. Their roles are discussed in Section~IV.
\subsection{Combined physics-informed and data-driven objective}\label{Combined-physics-informed-and-data-driven-objective}
The final optimization framework combines complementary quantitative metrics within a single objective function. The physics driven optimization evaluates source sensitivity and reference simulated MFM preservation, while the data driven optimization evaluates signal reconstruction and spatial features. The combined formulation is
\begin{equation}
\label{eq:27}
\fitmath{J_{combined}=J_{physics}+J_{data}}
\end{equation}
The quantities \(J_{\mathrm{physics}}\) and \(J_{\mathrm{data}}\) are the weighted objectives defined in \eqref{eq:16} and \eqref{eq:22}, respectively. Their outer coefficients in \eqref{eq:27} are both unity; the relative contribution of each block nevertheless depends on its internal weights and normalization.

The present study is combining information-based sensor-design criterion~\cite{ref25} with reference-field reconstruction criteria in a single weighted objective.
The three active physics-based terms were weighted 0.15, 0.70, and 0.15, respectively. 
For the data-driven configuration, the weights assigned to correlation, amplitude, pole position, gradient preservation, diversity, unweighted coverage, and weighted coverage were 0.51, 1.16, 0.80, 0.72, 0.74, 1.00, and 0.13, respectively. The remaining five terms such as dropout, pole separation, pole-pair orientation, trajectory error, and conditioning a weight of 0.05.
\subsection{Optimization algorithm}\label{Optimization-algorithm}
The sensor-layout optimization was performed using the sequential least squares quadratic programming (SLSQP) algorithm, a gradient-based constrained optimization method~\cite{ref32}. SLSQP iteratively updates the optimization variables to minimize a prescribed cost function while satisfying the specified geometric and feasibility constraints. In this study, the optimization variables comprised the \(2N\) lateral coordinates of \(N\) sensors. The objective function incorporated the relevant spatial Pearson correlation terms, Gaussian kernels, and the bilinearly interpolated lead field. 
At each iteration, SLSQP constructs and solves a local quadratic programming subproblem using a quadratic approximation of the objective function and linearized constraints. The resulting search direction is used to update the sensor coordinates, with step selection accounting for objective reduction and constraint satisfaction. This procedure is repeated until the convergence criteria are met or the prescribed iteration limit is reached.
In order to reduce dependence on the initial sensor configuration and assess the stability of the optimization, multiple initializations were employed. Each initialization defines a starting configuration of the sensor coordinates from which an independent SLSQP run is performed. The first search tier comprised 12 initializations, with a maximum of 1000 iterations per run and a convergence tolerance of \(\mathrm{f_{tol}}=10^{-6}\). If the predefined stability and consistency criteria were not satisfied, a second search tier was activated, comprising 16 additional initializations with a maximum of 1200 iterations per run.
\subsection{Evaluation}\label{Evaluation}
\begin{figure*}[!t]
    \centering
    \includegraphics[width=0.65\textwidth]
    {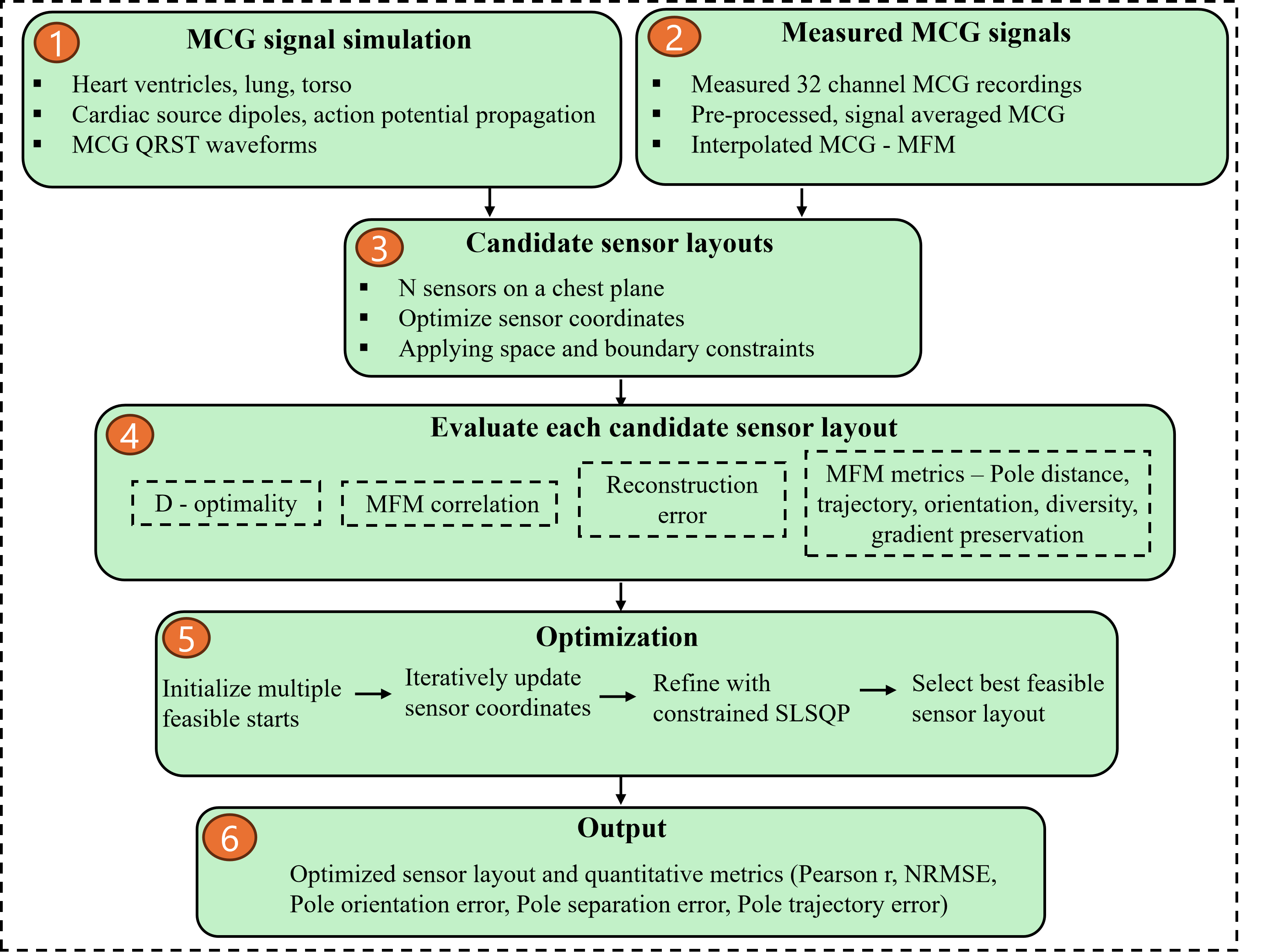}
    \caption{Overall workflow of all three methods}
    \label{fig:4}
\end{figure*}

The three optimization techniques are evaluated by comparing their reconstructed signals and magnetic-field maps with the 32-channel reference as described in Figure~\ref{fig:4}. The following five metrics are used to
quantify reconstruction performance after the optimization training:
\begin{enumerate}[label=(\roman*)]
\item {Pearson correlation ($r$):}
It measures similarity between the spatial distributions of the reference and reconstructed MFMs by calculating Pearson correlation across the reference positions at each valid timestamp and averaging across valid timestamps~\cite{ref20}.
\item {Normalized root-mean-square error (NRMSE):}
It quantifies the reconstruction error relative to the
reference signal strength. It is calculated as the square
root of the total squared reconstruction error divided by
the total squared reference signal across all sensor
positions and time points.
\item {Pole separation error (mm):}
It quantifies the difference in the separation distance
between the positive and negative magnetic field-lobe
centroids in the reconstructed maps relative to the
reference maps.
\item {Pole orientation error (deg):}
It measures the mean absolute angular difference between
the reference and reconstructed positive-to-negative pole
directions. The angular differences are wrapped to the
interval $[-180^\circ,180^\circ]$ before calculating their
absolute values.
\item {Pole trajectory error (mm):}
It evaluates the temporal consistency of the reconstructed
pole positions by calculating the mean Euclidean distance
between corresponding reference and reconstructed pole
centroids at matching time points, averaged across both
positive and negative polarities and all valid time instants.
\end{enumerate}
In the present study, the acceptance thresholds were kept as Pearson correlation
\(r \geq 0.99\), \(\mathrm{NRMSE} \leq 0.025\), pole separation
error \(<7\,\mathrm{mm}\), pole orientation error \(<5^\circ\),
and pole trajectory error \(<10\,\mathrm{mm}\) based on the research from earlier studies on coronary artery disease (CAD) diagnosis and ventricular repolarization \cite{ref33,ref34}. These thresholds
provide engineering criteria for selecting compact arrays with
adequate reconstruction fidelity and pole preservation; they
are not established clinical standards.

\section{Results}\label{Results}
\subsection{Physics driven optimization }\label{physics-drivenoptimization}
\begin{figure*}[t!]
    \centering
    \begin{subfigure}[t]{0.48\textwidth}
        \centering
        \includegraphics[width=\textwidth]{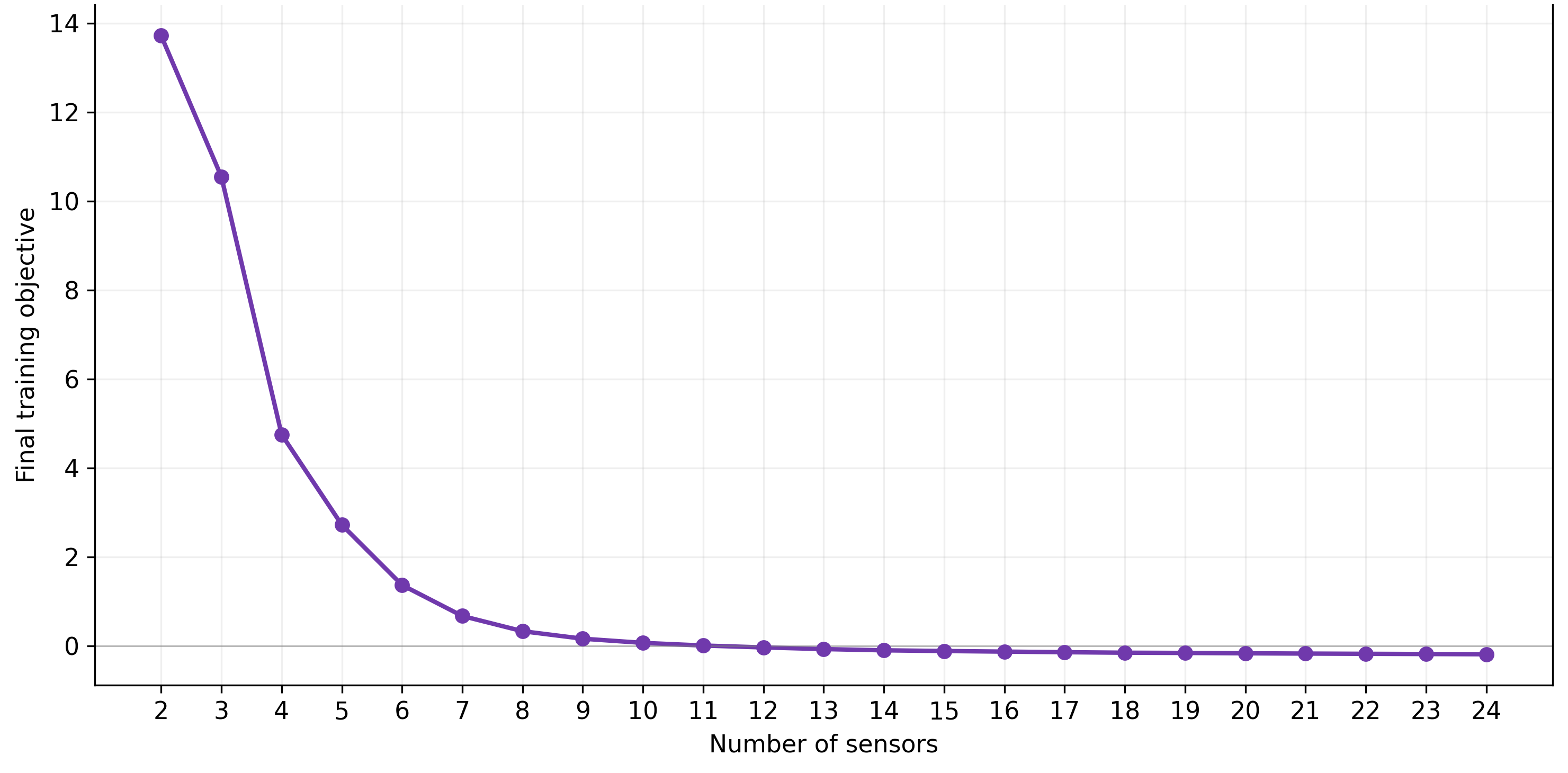}
        \caption{}
        \label{fig:physics_objective}
    \end{subfigure}
    \hfill
    \begin{subfigure}[t]{0.48\textwidth}
        \centering
        \includegraphics[width=\textwidth]{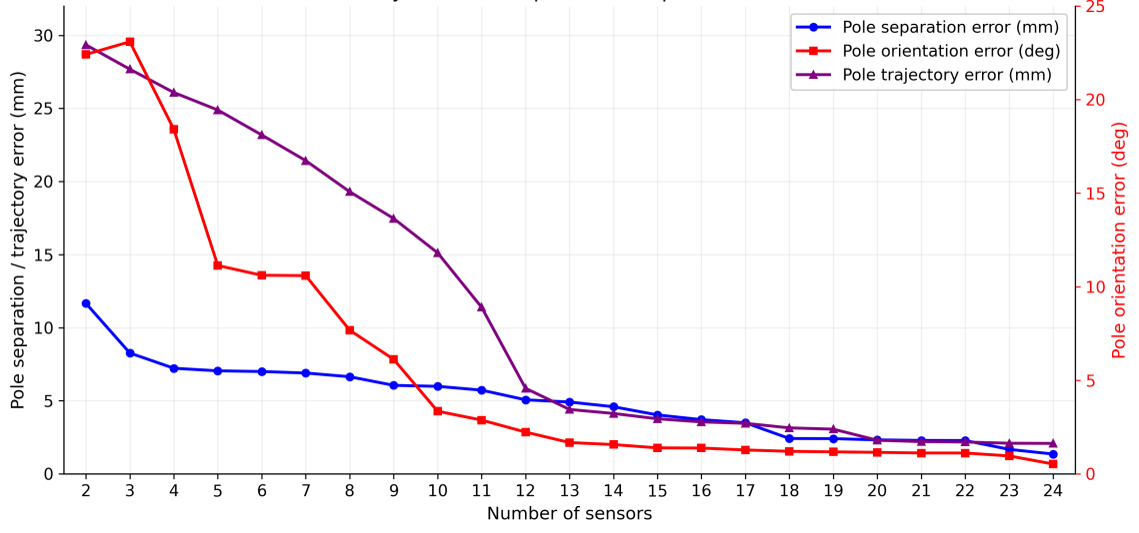}
        \caption{}
        \label{fig:pole_errors}
    \end{subfigure}
    \vspace{0.1cm}
    \begin{subfigure}[t]{0.48\textwidth}
        \centering
        \includegraphics[width=\textwidth]{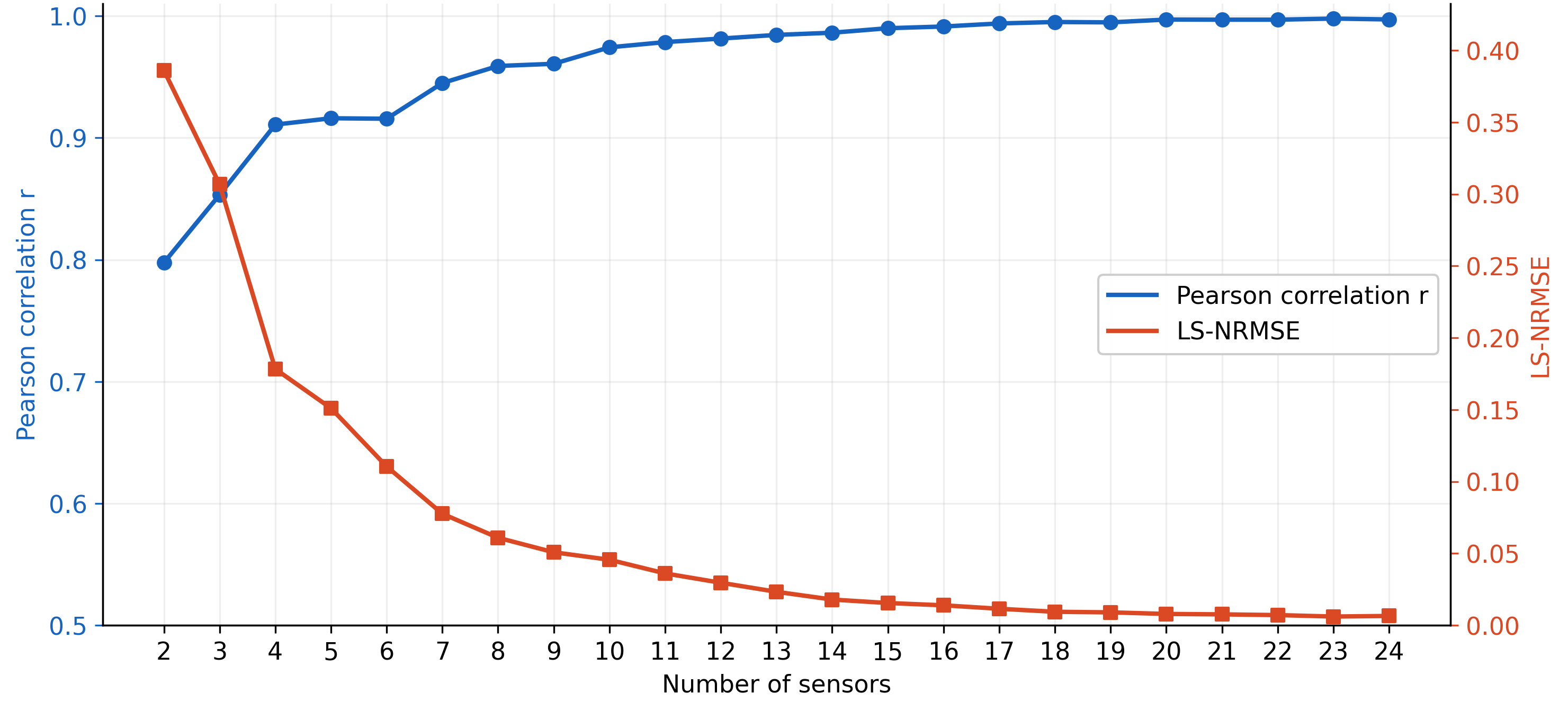}
        \caption{}
        \label{fig:reconstruction_metrics}
    \end{subfigure}
    \hfill
    \begin{subfigure}[t]{0.48\textwidth}
        \centering
        \includegraphics[width=\textwidth]{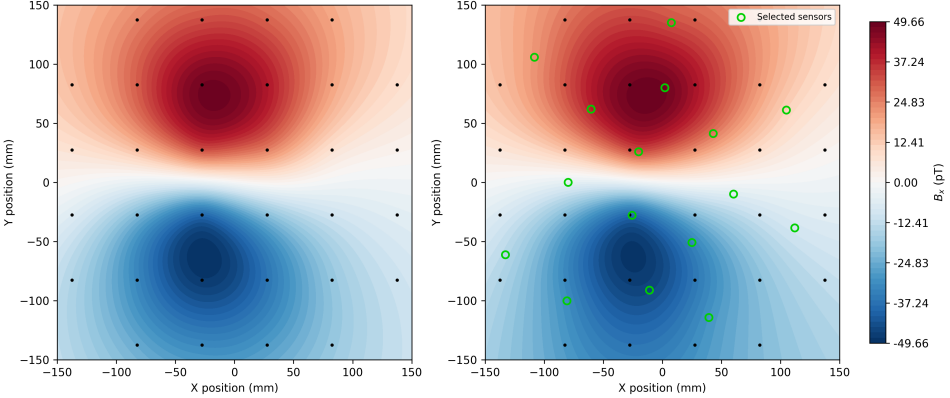}
        \caption{}
        \label{fig:reconstructed_MFM_phy}
    \end{subfigure}
    \caption{Performance of the physics-driven sensor optimization
    for different sensor counts: (a) final training objective;
    (b) pole separation, orientation, and trajectory errors;
    (c) Pearson correlation and LS-NRMSE;
    (d) T peak  Reference MFM vs Reconstructed  MFM for N = 16.}
    \label{fig:physics_driven_optimization}
\end{figure*}

In the physics-driven optimization, the final training objective decreases rapidly between two and seven sensors, followed by more gradual reductions up to sixteen sensors, as shown in Figure~\ref{fig:physics_driven_optimization}(\subref{fig:physics_objective}). This steep initial reduction indicates that additional sensing locations provide greater benefits to small arrays, with progressively diminishing improvements as the sensor count increases. Negative objective values are consistent with the negative log-determinant component. 
A similar trend is observed in the simulated MFM pole error metrics, which decrease with increasing sensor count, indicating improved preservation of pole spacing, orientation, and movement throughout the cardiac cycle, as shown in Figure~\ref{fig:physics_driven_optimization}(\subref{fig:pole_errors}). Beyond fourteen sensors, all three pole metrics exhibit minimal variation, suggesting that the reconstructed MFM patterns have largely stabilized. Pearson correlation and LS-NRMSE follow the same trend, approaching saturation beyond sixteen sensors at 0.9914 and 0.0139, respectively, as shown in Figure~\ref{fig:physics_driven_optimization}(\subref{fig:reconstruction_metrics}). Considering the high-fidelity criteria of Pearson correlation $\geq 0.99$ and LS-NRMSE $\leq 0.025$, the minimum effective physics-driven configuration is identified as sixteen sensors, providing a compact array with high reconstruction fidelity.
The corresponding sensor layouts also reveal how spatial coverage evolves with increasing array size. Smaller array configurations concentrate sensors around the dominant magnetic poles and the intervening region, whereas larger arrays distribute sensors around the lobe boundaries and toward the periphery, capturing more of the spatial variation as shown in Figure~\ref{fig:physics_driven_optimization}(\subref{fig:reconstructed_MFM_phy}). This progression indicates a transition from capturing the dominant bipolar field structure to resolving its spatial extent and finer variations.
\subsection{Data driven optimization}\label{data-driven-optimization-result}
\begin{figure*}[t!]
    \centering
    \begin{subfigure}[t]{0.48\textwidth}
        \centering
        \includegraphics[width=\textwidth]{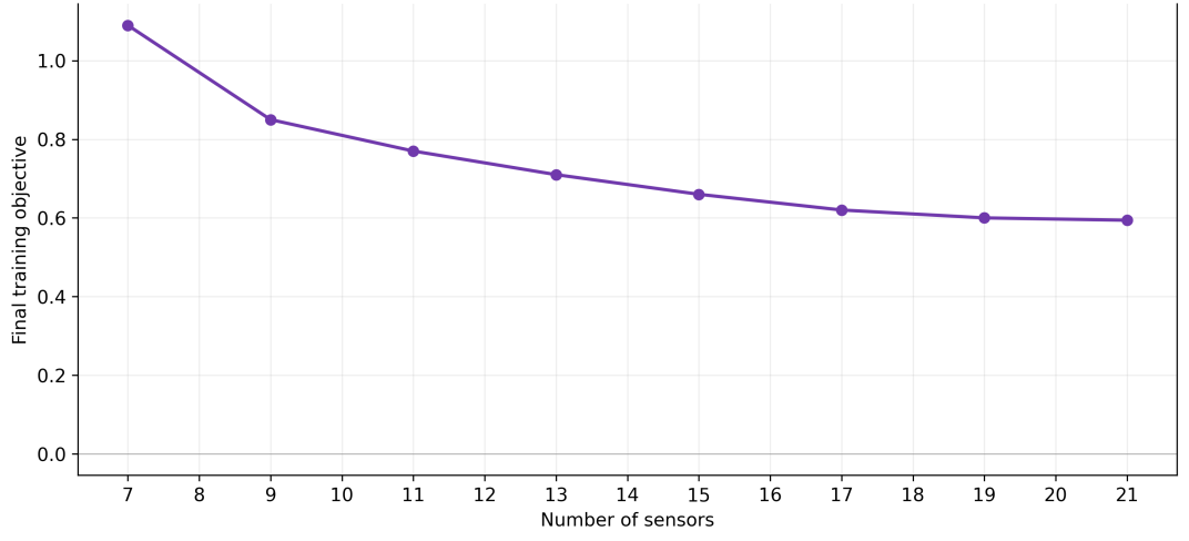}
        \caption{}
        \label{fig:data_objective}
    \end{subfigure}
    \hfill
    \begin{subfigure}[t]{0.48\textwidth}
        \centering
        \includegraphics[width=\textwidth]{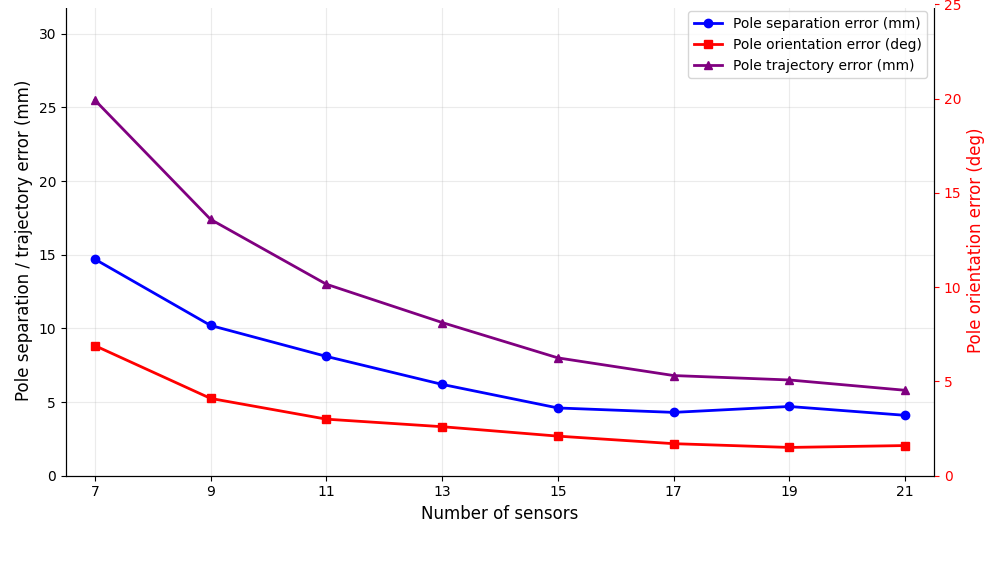}
        \caption{}
        \label{fig:pole_errors_data}
    \end{subfigure}
    \vspace{0.1cm}
    \begin{subfigure}[t]{0.48\textwidth}
        \centering
        \includegraphics[width=\textwidth]{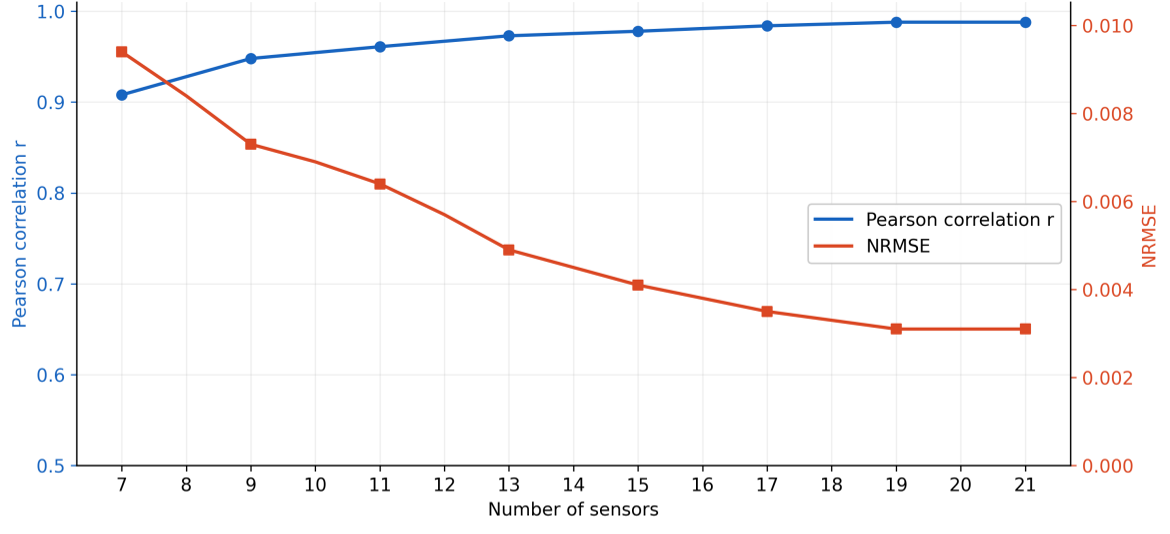}
        \caption{}
        \label{fig:reconstruction_metrics_data}
    \end{subfigure}
    \hfill
    \begin{subfigure}[t]{0.48\textwidth}
        \centering
        \includegraphics[width=\textwidth]{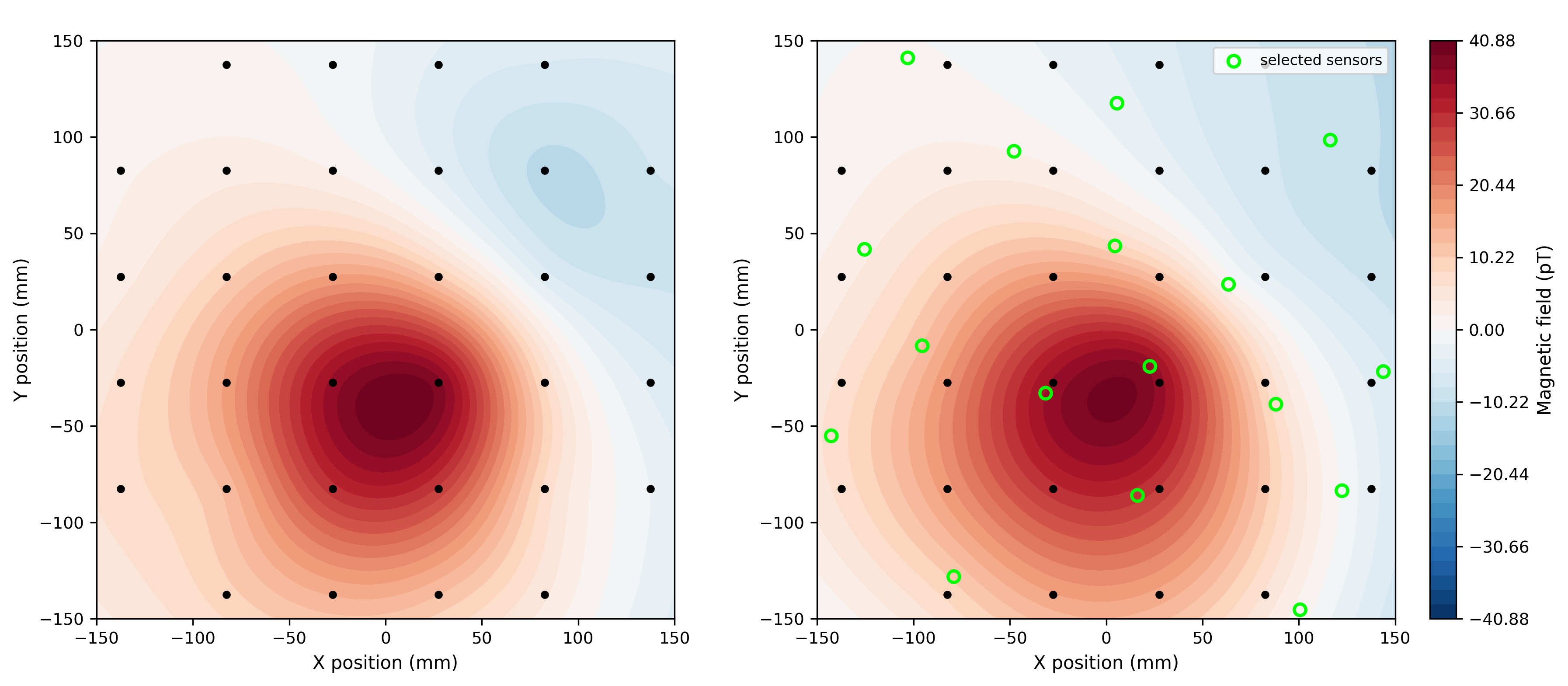}
        \caption{}
        \label{fig:reconstructed_MFM_data}
    \end{subfigure}
    \caption{Performance of the data driven sensor optimization
    for different sensor counts: (a) final data training objective;
    (b) pole separation, orientation, and trajectory errors;
    (c) Pearson correlation and LS-NRMSE; 
    (d) T peak Reference MFM vs Reconstructed MFM for N =17 sensors.}
    \label{fig:data_driven_optimization}
\end{figure*}

The final training objective in the data-driven optimization decreases from approximately 1.090 at seven sensors to 0.620 at seventeen sensors, accounting for approximately \(94.8 \%\) of the total objective reduction as evident in Figure~\ref{fig:data_driven_optimization}(\subref{fig:data_objective}). Beyond seventeen sensors, adding four more sensors produces only marginal improvements, with a correlation increase of 0.004 and an NRMSE reduction of 0.0004.
The three pole error metrics also exhibit an overall decreasing trend with increasing sensor count, followed by a flattening of the curves beyond seventeen sensors, with only minor fluctuations in pole separation and orientation errors as shown in Figure~\ref{fig:data_driven_optimization}(\subref{fig:pole_errors_data}).
Similarly, Pearson correlation increases from 0.9080 at seven sensors to 0.9840 at seventeen sensors, while NRMSE decreases from approximately 0.0094 to 0.0035 as shown in Figure~\ref{fig:data_driven_optimization}(\subref{fig:reconstruction_metrics_data}). These trends indicate improved preservation of the reference MFM with increasing sensor count, followed by diminishing improvements beyond seventeen sensors. Although the correlation remains below the predefined threshold of 0.99, the relatively small gains beyond seventeen sensors support this configuration as a practical compromise between reconstruction fidelity and array size.
Unlike the physics-driven layouts, which initially concentrate around the dominant magnetic poles, the data-driven configurations exhibit broader spatial coverage even at lower sensor counts. Sensors are distributed across the dominant positive lobe, field transition regions, and outer boundaries. With increasing sensor count, the additional locations progressively fill the spatial gaps while maintaining coverage of both positive and negative field regions. This distribution suggests that optimization against experimentally measured MFMs favours the preservation of broader spatial field characteristics rather than concentrating primarily on the dominant field extrema as shown in Figure~\ref{fig:data_driven_optimization}(\subref{fig:reconstructed_MFM_data}).
\subsection{Combined parameters optimization results}\label{combined-parameters-optimization-results}
\begin{figure*}[t!]
    \centering
    \begin{subfigure}[t]{0.48\textwidth}
        \centering
        \includegraphics[width=\textwidth]{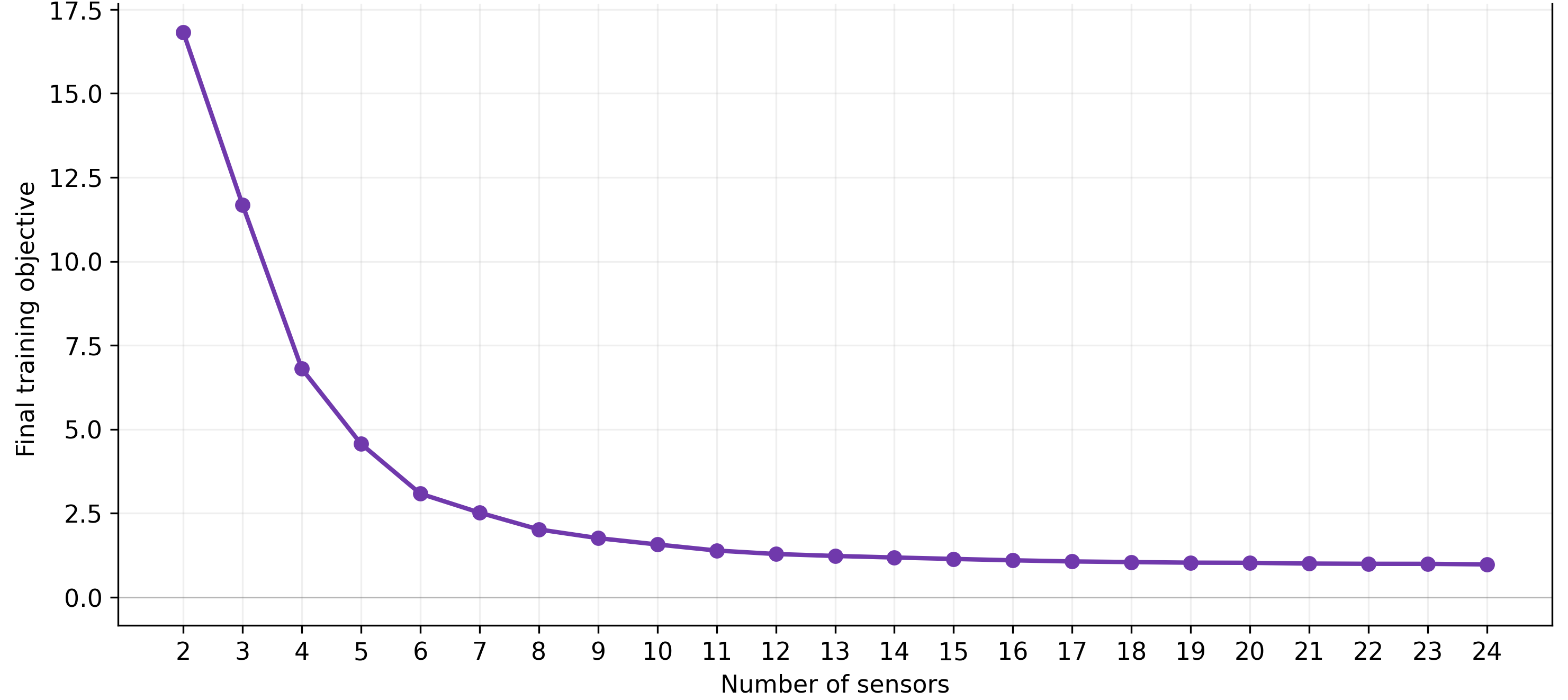}
        \caption{}
        \label{fig:combined_objective}
    \end{subfigure}
    \hfill
    \begin{subfigure}[t]{0.48\textwidth}
        \centering
        \includegraphics[width=\textwidth]{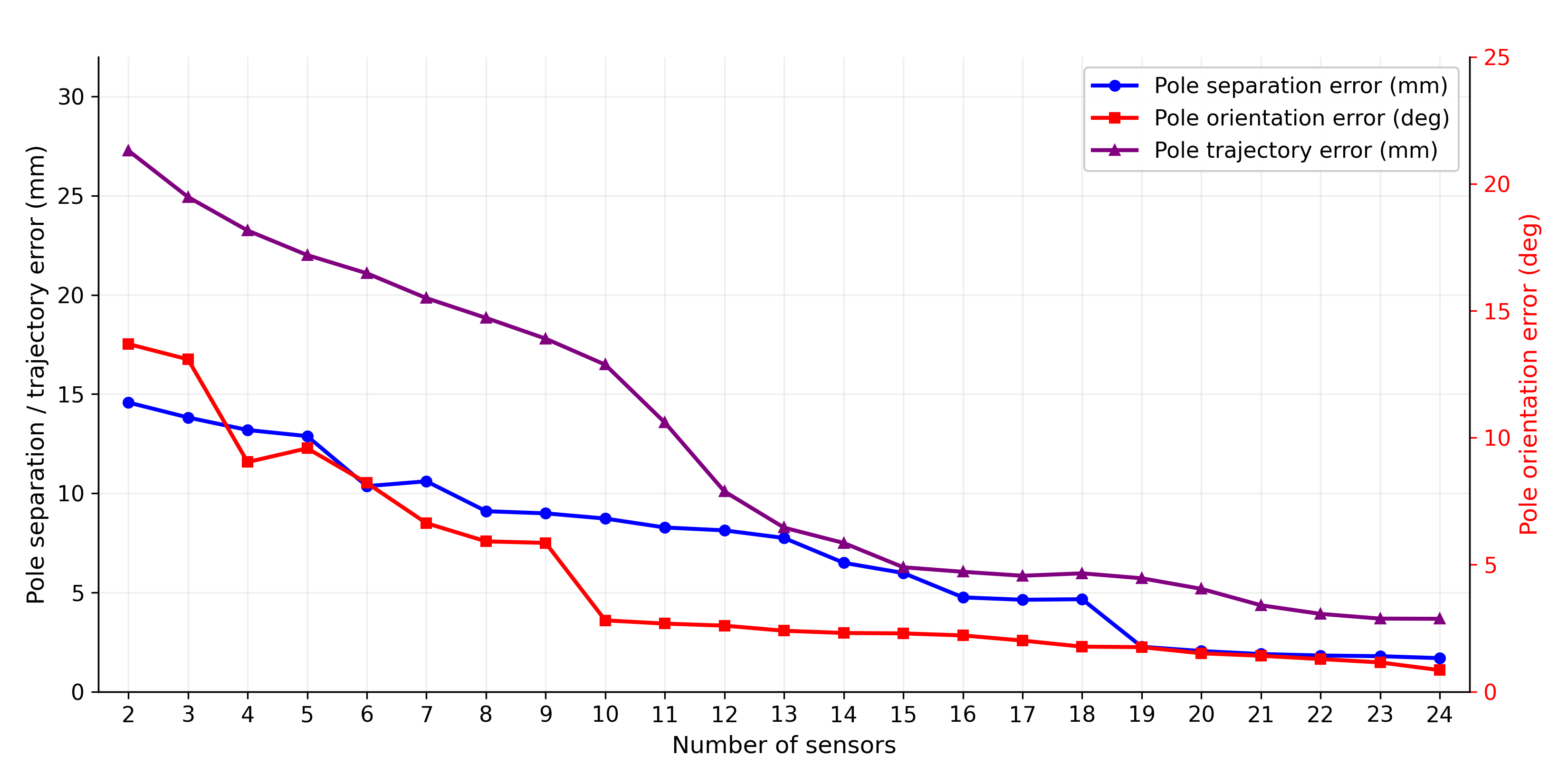}
        \caption{}
        \label{fig:pole_errors_combined}
    \end{subfigure}
    \vspace{0.1cm}
    \begin{subfigure}[t]{0.48\textwidth}
        \centering
        \includegraphics[width=\textwidth]{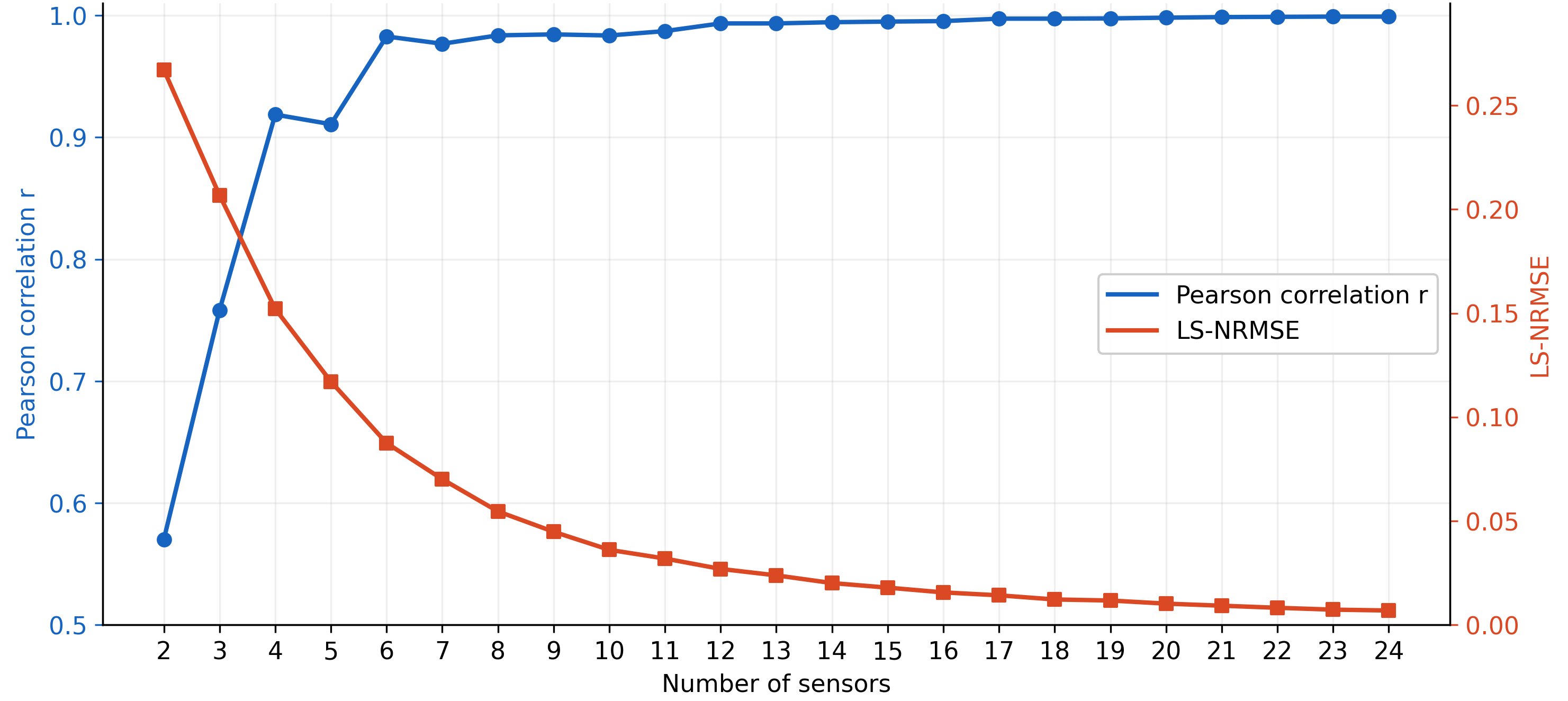}
        \caption{}
        \label{fig:reconstruction_metrics_combined}
    \end{subfigure}
    \hfill
    \begin{subfigure}[t]{0.51\textwidth}
        \centering
        \includegraphics[width=\textwidth]{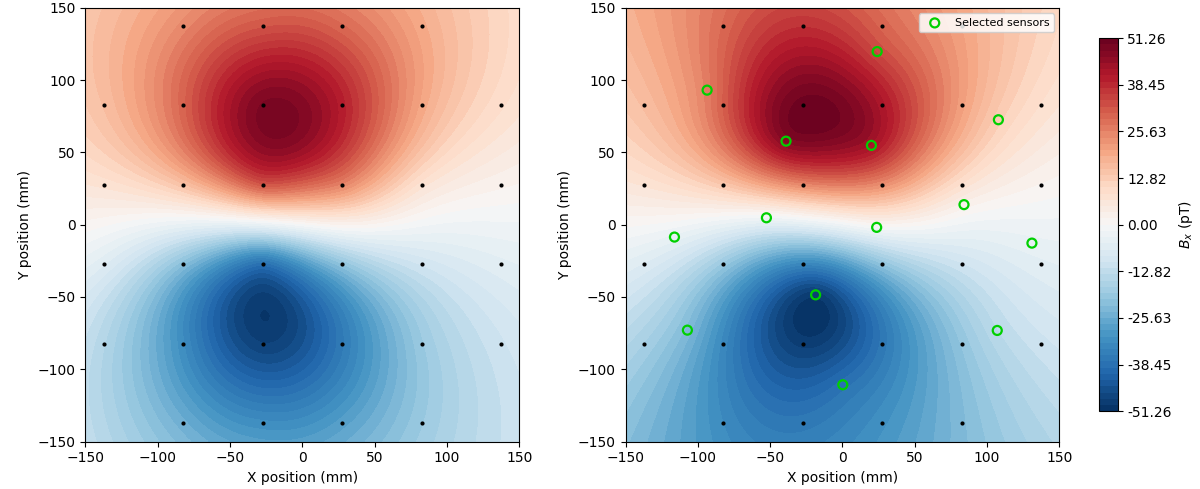}
        \caption{}
        \label{fig:reconstructed_MFM_combined}
    \end{subfigure}
    \vspace{0.1cm}
    \begin{subfigure}[t]{0.3\textwidth}
        \centering
        \includegraphics[width=\textwidth]{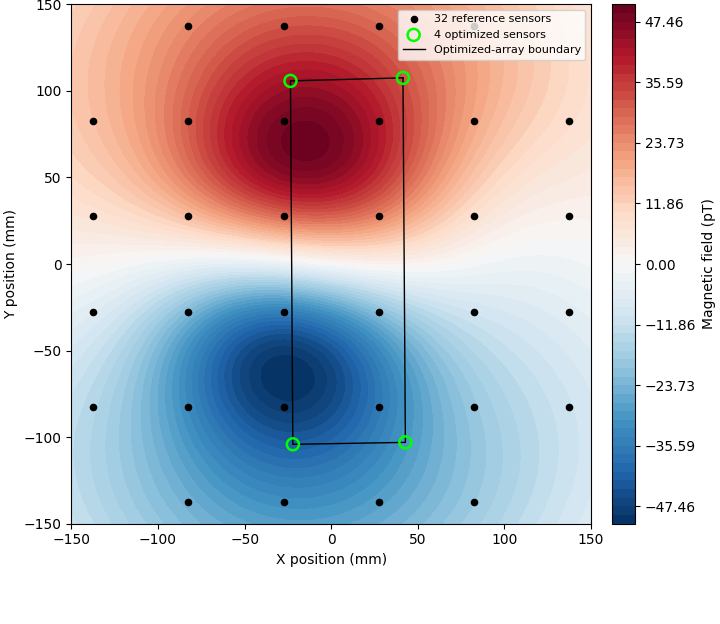}
        \caption{}
        \label{fig:MFM_on_ref04}
    \end{subfigure}
    \hfill
    \begin{subfigure}[t]{0.3\textwidth}
        \centering
        \includegraphics[width=\textwidth]{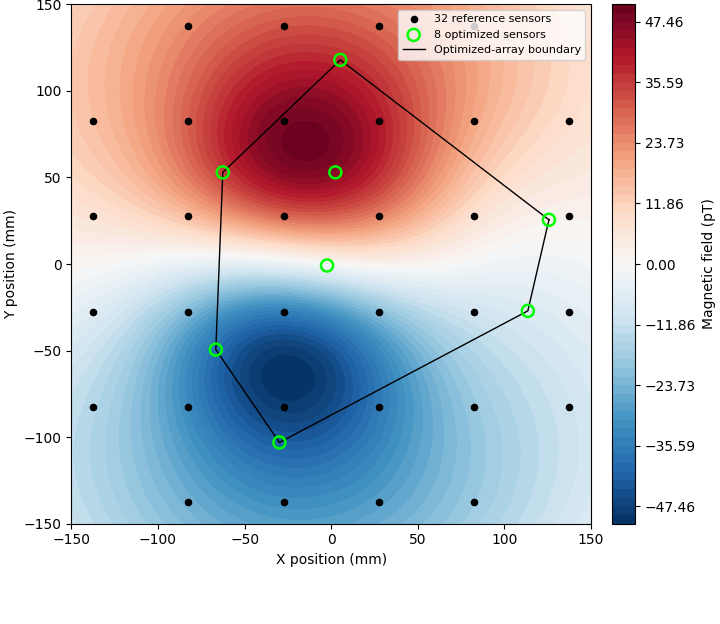}
        \caption{}
        \label{fig:MFM_on_ref08}
    \end{subfigure}
    \hfill
    \begin{subfigure}[t]{0.3\textwidth}
        \centering
        \includegraphics[width=\textwidth]{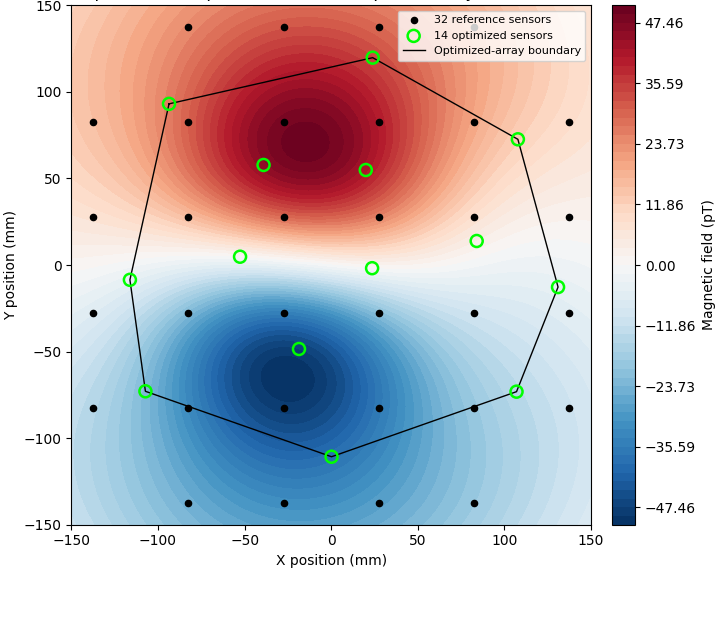}
        \caption{}
        \label{fig:MFM_on_ref14}
    \end{subfigure}
    \caption{Performance of the combined parameters sensor optimization
    for different sensor counts: (a) final combined training objective;
    (b) pole separation, orientation, and trajectory errors;
    (c) Pearson correlation and LS-NRMSE; 
    (d) T peak Reference MFM vs Reconstructed MFM for N = 14;
    Optimized sensor layout placed on reference at: (e) N=4; (f) N=8; (g) N=14 sensors respectively.}
    \label{fig:combined_objective_optimization}
\end{figure*}

The combined optimization exhibits progressive improvements in the training objective and pole error metrics with increasing sensor count, as shown in Figure~\ref{fig:combined_objective_optimization}(\subref{fig:combined_objective}) and Figure~\ref{fig:combined_objective_optimization}(\subref{fig:pole_errors_combined}). 
The observed variations at intermediate sensor counts reflect changes in the optimized configurations under the joint objective, which balances the preservation of simulated cardiac source information with the reconstruction of experimentally measured MFMs. This behaviour indicates that the combined parameters formulation incorporates the complementary requirements of the physics-driven and data-driven approaches. 
Importantly, the combined parameters formulation achieves the predefined reconstruction criteria of Pearson correlation $\geq 0.99$ and LS-NRMSE $\leq 0.025$ with fourteen sensors, with both criteria maintained at all larger evaluated sensor counts as shown in Figure~\ref{fig:combined_objective_optimization}(\subref{fig:reconstruction_metrics_combined}). Compared with the sixteen-sensor physics-driven configuration, this represents a reduction of two sensors 12.5 \(\%\) while satisfying the same reconstruction requirements, demonstrating improved sensor efficiency through the joint optimization .
The spatial distribution of the optimized configurations further illustrates the influence of the combined objective, as illustrated in Figure~\ref{fig:combined_objective_optimization}(\subref{fig:reconstructed_MFM_combined}). The four-sensor arrangement spans the dominant positive and negative magnetic-field lobes, as shown in Figure~\ref{fig:combined_objective_optimization}(\subref{fig:MFM_on_ref04}). With eight sensors, the configuration extends towards the right peripheral region while introducing additional sampling within the positive lobe and polarity-transition region, as shown in Figure~\ref{fig:combined_objective_optimization}(\subref{fig:MFM_on_ref08}). At fourteen sensors, the array provides broader coverage across both lateral regions, with additional locations distributed around the magnetic lobes and central transition region, as seen in  Figure~\ref{fig:combined_objective_optimization}(\subref{fig:MFM_on_ref14}). This distribution reflects a balance between sampling the dominant magnetic-field extrema and capturing the spatial variations across the broader field.
\section{Discussion}\label{Discussion}
This study optimized the frontal-plane sensor configuration for MCG by integrating a physics-driven approach based on simulated cardiac signals using the LFM with a data-driven approach using experimentally acquired MCG signals, considering a 32-channel array as the reference configuration. The results demonstrate that a reduced, non-uniform sensor arrangement can preserve the essential MCG signal information with minimal loss compared with the full, uniformly distributed 32-channel array~\cite{ref11,ref19,ref20}. A reduced subset of 14 sensors was sufficient to reconstruct the magnetic field characteristics of the 32-sensor reference, and the optimized layouts retained roughly half of the original array, distributed non-uniformly inline with the results of the earlier published study~\cite{ref20}. Since 14 of 32 is about 43 \(\%\) of the original array, the result shows that a substantial part of the reference array is redundant. 
Each of the three optimization methods yielded a different sensor layout and optimal sensor count. This may be attributed to the existence of multiple sensor subsets that provide comparable field reconstruction performance, owing to the strong spatial correlation between neighbouring channels in an MCG array. Furthermore, differences in the objective functions and their relative weighting of signal characteristics may lead to convergence towards different local optima with comparable performance. However, once the target number of sensors was reached, the resulting solutions exhibited minimal variation across the three methods, suggesting that the optimization had reached a near optimal solution. The reason could be the usage of 15 different cost functions each looking for different characteristics of the MCG signals for the optimization. In the combined parameters method, the weights were chosen to favour the spatial characteristics of the MCG signal: D-optimality received a lower weight and the pole metrics a higher one. LFM based optimization alone tends to push sensors away from the heart, because the lead field is better conditioned there, and the resulting loss of signal strength degrades the measurement quality. Assigning greater weight to the pole metrics encourages the optimization to retain sensor positions that capture the dipolar field patterns, thereby overcomes this issue~\cite{ref11,ref18}. 
The physics-based and data-driven approaches were combined because each has a weakness the other covers. LFM based optimization cannot capture all the information in measured MCG signals, and its optimal array depends on the anatomy used, so an array optimized for one geometry need not be optimal for another. A data-driven optimization uses real recordings from several subjects for training and evaluation, which brings in this missing information. Its weakness is that it depends on the available data, and the number of subjects was limited. Combining the two gives a balance between physical reasoning and empirical evidence. 
Beyond field reconstruction and pole preservation, the data‑driven objective added penalties for sensor redundancy, spatial coverage, sensor dropout and numerical conditioning. The diversity penalty pushed sensors with waveforms apart urging complementary measurements~\cite{ref20}. Conventional coverage penalty made large gaps between evaluation points and their sensor costly whereas activity‑weighted coverage highlighted areas, with more magnetic‑field change. The dropout penalty looked at how sensitive reconstruction was when a single sensor was removed and the conditioning penalty warned against unstable thin‑plate‑spline reconstruction systems. All these terms pushed sensor layouts to balance reconstruction fidelity with coverage and robustness.
Earlier works using the LFM-CN as the metric found that sensors concentrate on the regions of strong field gradient and the boundary of the search volume~\cite{ref11,ref14}. In the present study, the sensors first preserved the dipole maximum and minimum, and only then the regions of high gradient. D-optimality therefore appears to be a more useful measure of LFM sensitivity than the condition number. One possible explanation is that the condition number measures only how well-conditioned the inverse problem is. It does not measure how well the sensors represent the shape of the measured field. Sensors chosen by CN can therefore sit at the edge of the volume, where they are well-conditioned but far from the heart and weak in signal. D-optimality, which maximizes the determinant of the information matrix, rewards the total information content of the selected sensors and so appears to be a more useful measure of LFM sensitivity~\cite{ref25}. A second possible reason for the difference lies in the source model. Earlier studies built the LFM and generated the MFM under a reduced source-space assumption, which cannot represent the simultaneous activation of dipoles at multiple locations produced by the propagating action potential wavefront. In the present work, all dipoles in the ventricular myocardium were included in the simulated MCG-MFM and in the resulting LFM, so the physiological information of the distributed activation is retained. The preservation of the dipole extrema is clinically relevant. Pole angle and pole pattern of the MFM are independent biomarkers for early detection of CAD~\cite{ref3,ref9}. An array that preserves these features can therefore support diagnostic interpretation as well as signal reconstruction, which makes the optimized layout more useful than one chosen on mathematical criteria alone.
The simulations were performed using an intersensor spacing of 55 mm to match the existing experimentally recorded MCG data. During sensor position optimization, an intersensor spacing constraint of 54 – 65 mm was imposed to identify practically feasible sensor configurations for room temperature MCG systems, since it limits crosstalk between neighbouring sensors and leaves room for the sensor housing. Studies based on the spatial Nyquist sampling theorem report that a 30 mm inter-sensor spacing captures the entire cardiac electrophysiological cycle for both simulated and real MCG data. Additionally, prominent signal amplitude waveforms of real MCG signals can still be obtained at spacings up to 65 mm~\cite{ref35}. Other works suggested that the needed spacing seems to depend on the ratio of source – sensor  distance to inter-sensor distance: the farther the sensors are from the source, the smoother the field they record, and the wider the spacing that can be tolerated~\cite{ref36}. A spacing of 54 - 65 mm lies below the maximum constraint and is sufficient to recover the information contained in the ventricular QRS and T-wave activation wavefronts.
Subject-specific sensor optimization is important for accounting for anatomical variations and improving the accuracy of cardiac field measurements. However, implementing individualized sensor configurations would require a customized array for each subject, limiting its practical applicability. A common optimized sensor configuration that provides consistent performance across subjects would therefore be more suitable for standardizing MCG data acquisition and clinical interpretation. Considering the variations in sensor technologies, sensitivities, and measurement environments, the proposed framework provides a step towards this goal by identifying reduced sensor configurations that preserve essential cardiac source information and spatial field characteristics, potentially supporting consistent source localization and diagnosis across different MCG systems~\cite{ref37}.
\subsection{Limitations and future work}\label{Limitations-and-future-work}
Since real MCG data were available only for 32 sensors, the simulation of MCG signals and resulting LFM were constructed for the same 32 sensors. The optimization was therefore limited to a reduced subset of this reference, and the result may differ from a universal optimum for total number of sensor channels. The present study did not consider the effects of sensor arrays on lateral and posterior side of the torso. Even though the lateral and posterior side has been studied for ischemia and atrial activity, it suffers from practical difficulty in keeping the sensors and its contribution for clinical diagnosis remains poorly evidenced~\cite{ref38,ref39}. The physics-based approach used a geometry with piecewise-homogeneous conductivities, which simplifies the torso. Anatomical models from a diverse cohort of subjects should be incorporated to improve the generalizability of the physics-based approach.
The data-driven analysis used only a few datasets, and subjects with a clear dipolar MFM and matching R and T wave peak MFMs. This selection improves the consistency of the training data but restricts the generality of the result. A universal array would need to be optimized over a much larger set of subjects covering different age groups and cardiac conditions. Neither the simulations nor the measured data included cardiac anomalies such as arrhythmia or CAD. The optimal positions may vary with the type of abnormality and its severity, so the array needs to be tested on a range of disorders before clinical use~\cite{ref12,ref20}. The influence of noise on the individual cost functions, and consequently on the chosen sensor array, was not studied. Noise could change which sensors appear most valuable, especially for weak signals such as the T wave. In future, the possible extension could be addressing these limitations by optimizing sensor array for larger and more diverse datasets and anatomies, including pathological cases and measuring at several inter-sensor spacings with different noise added to signals so that proposed optimization methodology can also be applicable for unshielded MCG systems.
\section{Conclusion}\label{Conclusion}
The present work focused on developing a hybrid framework to optimize the sensor array of a multichannel MCG system. It combines a physics-driven objective (D-optimality, spatial correlation of the simulated T wave, and least-squares error) with a data-driven objective based on measured MFMs and their pole-geometry metrics. Each framework was used to find the sensor count and continuous frontal-plane positions that preserve the information of a 32-channel reference array. The proposed hybrid approach identified 14 sensor positions as the optimized one (Pearson correlation = 0.9945, NRMSE = 0.0201, pole separation error = 6.4953 (mm) pole orientation error = 2.3107 (deg) and pole trajectory error = 7.4952 (mm)) compared to the reference. The proposed framework offers a practical applicability of optimized sensor arrangement for room temperature cost effective, flexible, portable MCG systems. However, further validation is needed for direct translation to existing clinical MCG setups.
\section{Acknowledgement}\label{Acknowledgement}
The authors acknowledge the financial support and funding received from the Qmet Tech Foundation, Department of Science and Technology (DST) under the National Quantum Mission (NQM) of Government of India, for performing this study.
\section{Data availability statement}\label{Data availability statement}
The data that support the findings of this study are available from the corresponding author upon reasonable request.

\end{document}